\documentclass[11pt,a4paper]{article}

\usepackage{my-j}

\usepackage{amssymb}
\usepackage{amsfonts}
\usepackage{graphicx}
\usepackage{epstopdf}
\usepackage{dcolumn}
\usepackage{amsmath}
\usepackage{latexsym,bm}
\usepackage{amsthm}
\usepackage{slashed}
\usepackage{float}
\usepackage{color}
\usepackage{url}

\def \be {\begin{equation}}
\def \ee {\end{equation}}
\def \bsp {\begin{split}}
\def \esp {\end{split}}
\def \bea {\begin{eqnarray}}
\def \eea {\end{eqnarray}}

\def\mc{\mathcal}

\def\P{\mathbb{P}}

\def\fset{\mc{F}}
\def\gset{\mc{G}}

\title{Scanning the skeleton of the 4D F-theory landscape}

\author[a]{Washington Taylor, Yi-Nan Wang}

\affiliation[a]{Center for Theoretical Physics,\\Department of Physics\\Massachusetts Institute of Technology\\77 Massachusetts Avenue\\Cambridge, MA 02139, USA}

\emailAdd{wati\ {\rm at}\ mit.edu, wangyn\ {\rm at}\ mit.edu}

\preprint{\today \hspace*{0.1in} MIT-CTP-4936}

\abstract{Using a one-way Monte Carlo algorithm from several different
  starting points, we get an approximation to the distribution of
  toric threefold bases that can be used in four-dimensional F-theory
  compactification. 
We separate the threefold bases into
  ``resolvable'' ones where the Weierstrass polynomials $(f,g)$ can
  vanish to order (4,6) or higher on codimension-two loci and the
  ``good'' bases where these (4,6) loci are not allowed. 
A simple estimate suggests that the number of distinct resolvable base
geometries exceeds
$10^{3000}$, with over $10^{250}$ ``good'' bases, though the actual
numbers are likely much larger.
We find that
  the good bases are concentrated at specific ``end points'' with
  special isolated values of $h^{1,1}$ that are bigger than
  1,000. 
These end point bases
give Calabi-Yau fourfolds with specific Hodge numbers mirror to elliptic
fibrations over simple threefolds.
The non-Higgsable gauge groups on the end point bases are
  almost entirely made of products of $E_8$, $F_4$, $G_2$ and
  SU(2). Nonetheless, we find a large class of good bases with a
  single non-Higgsable SU(3). Moreover, by randomly contracting the
  end point bases, we find many resolvable bases with $h^{1,1}(B)\sim
  50$-200 that cannot be contracted to another smooth threefold
  base. }


\begin{document}

\maketitle

\section{Introduction}

F-theory~\cite{Vafa-F-theory, Morrison-Vafa-I, Morrison-Vafa-II} is a
useful geometric framework that describes a large class of strongly
coupled IIB superstring compactifications. The geometric data in an
F-theory construction consist of a complex $d$-dimensional manifold
$B$ and an elliptic Calabi-Yau fibration $X$ over $B$; each choice of
$B, X$ leads
to a certain $(10-2d)$-dimensional low-energy physics.

A grand classification program of all the possible F-theory
compactifications to $\mathbb{R}^{9-2d,1}$ can be carried out in the
following three steps:

(1) Classify
up to
isomorphism all the $d$-dimensional base manifolds $B$ that can
support an elliptic Calabi-Yau $(d+1)$-fold. For a large fraction of bases, the generic elliptic
fibration $X$ over $B$ has singularities corresponding to 7-branes
carrying gauge groups. These gauge groups are called non-Higgsable
gauge groups~\cite{clusters,4d-NHC}, and are a characteristic
feature of the
base geometry $B$ itself.  For $d=2$, the base manifolds have been
almost completely
classified~\cite{mt-toric,Hodge,Martini-WT,non-toric}.

(2) Classify all the topologically distinct (non-generic) elliptic
fibrations over $B$. The Weierstrass model of an elliptic fibration
$X'$ is usually singular, and  can be resolved to another
Calabi-Yau manifold $X$. We are actually interested in classifying the different $X$
up to isomorphism. In the physics language, different fibrations
correspond to different gauge groups and matter
content~\cite{Bershadsky-all, Katz-Vafa, Morrison-sn,
  Grassi-Morrison-2, mt-singularities, Johnson:2016qar}. For a
non-generic elliptic fibration, the geometric gauge group is always
larger than the non-Higgsable gauge group on $B$, and the number of
complex structure moduli generally decreases.

(3) Classify other non-geometric information that affects the low
energy physics, such as $G_4$ flux~\cite{Grana:2005jc, Douglas:2006es,
  Denef-F-theory, dgkt, Acharya:2006zw, Braun-Watari, Watari} or
T-brane
structures~\cite{Donagi:2003hh,Cecotti:2010bp,Anderson:2013rka}. The
$G_4$ flux has especially important impact  on the physics of 4D F-theory
compactifications, such as producing net chirality for matter and
generating the Gukov-Vafa-Witten superpotential~\cite{GVW}. But we
will not take it into account in this paper.

In this paper, we focus on the classification of complex threefold
bases, which is the foundation of this whole classification program
for 4D F-theory. We restrict ourselves to the set of 3D toric bases,
where we have good control on the birational transitions (blow
up/down) between one base and another. The special class of bases that
have the form of $\mathbb{P}^1$ bundles over toric surfaces was
studied in \cite{Halverson-WT}; this subset contains only $ \sim 10^5$
3D toric bases. A more general partial classification of toric
threefold bases was done in our previous work~\cite{MC}, where we
performed random sequences of blow up/downs starting from the base
$\mathbb{P}^3$. For technical reasons, we restricted ourselves to the
set of bases connected through transitions without any intermediate codimension-two (4,6) locus, which effectively
leaves out the bases with non-Higgsable $E_8$s. The estimated total number of bases in this restricted set is
about $10^{48}$. In a more recent work by Long, Halverson and
Sung \cite{Halverson:2017}, the more general set of toric bases
allowing codimension-two (4,6) loci is considered. a lower bound of the
total number of bases from blowing up $\mathbb{P}^3$ was estimated as
$4/3\times 2.96\times 10^{755}$. However, this is also a restricted
subset since they
 imposed fairly stringent bounds on the number and structure of
 allowed blow-ups to avoid
overcounting using their systematic approach.

In this work, we perform a sequence of random blow ups starting from a
fixed base, using a one-way version of the Monte Carlo approach in
\cite{MC}. Toric codimension-two (4,6) loci are allowed to appear on
the bases in this process. We call bases with such loci ``resolvable
bases'' in this paper.  From the point of view of the elliptic
Calabi-Yau fourfold, the codimension-two (4, 6) locus indicates a
non-flat fibration; a Calabi-Yau resolution can generally be achieved
by first blowing up the (4, 6) locus in the base and then resolving
further local Kodaira singularities in the usual way.  Physically, the
presence of a codimension-two (4, 6) locus indicates the presence of
an infinite family of massless states in the low-energy theory, and
the general consequences of this for physics in 4D are not fully
understood.  In the context of 6D F-theory, a (4, 6) point on the base
surface indicates the presence of a 6D (1,0) superconformal field
theory (SCFT) \cite{Seiberg,Heckman:2013pva,DelZotto:2014hpa,Heckman:2015bfa},
and blowing up the point in the base corresponds to taking the SCFT
onto the tensor branch.  We call a toric threefold generated in the
blow-up process a ``good base'' if it has no toric codimension-two (4,6)
locus, since no toric blow-up of the base is necessary for a maximal crepant
resolution\footnote{There may be some  (non-toric) codimension-two (4,6) loci
  on divisors with $E_8$ gauge group, but these loci can be blown up
  without changing the non-Higgsable gauge group structure}.

The set of  toric threefold base geometries  explored in this
paper and the earlier works \cite{MC, Halverson:2017} 
can be thought of as the ``skeleton'' of the landscape
of 4D ${\cal N} = 1$ F-theory vacua.   Specifically, there is a large
but finite
set of toric threefold bases that are connected to one another and to
simple toric bases like $\P^3$ by blow up and blow down transitions
with a simple toric description as adding or removing rays in the
toric fan.
\footnote{The fact that this set is finite is a corollary of the more
  general recent results that the number of threefold base geometries
  and the total number of elliptic Calabi-Yau fourfold geometries are
  finite (up to birational equivalence), and the number of distinct
  sets of Hodge numbers are therefore finite \cite{dc-s}.}  This large
connected moduli space forms arguably the most concrete mathematical
realization so far of a systematic description of a large connected
portion of the space of 4D ${\cal N} = 1$ supergravity theories. This
``skeleton'' is only a coarse picture of the full space of theories;
there are a variety of non-toric F-theory threefold bases possible, it
has not been proven that all allowed toric threefold bases are in the
connected set, and for each base geometry there can be many different
elliptic Calabi-Yau fourfolds with different physics, with a
potentially very large multiplicity of flux vacua for each CY
fourfold, separated by a superpotential that lifts most or all moduli
in the connected geometric moduli space.  Nevertheless, understanding
this skeleton of the full space of theories seems like a worthwhile
first step in obtaining a global understanding of the space of ${\cal
  N} = 1$ theories in 4D.  Furthermore, the analyses mentioned above
of the corresponding problem in 6D show that the roughly 65,000
``good'' toric base surfaces that support elliptic Calabi-Yau
threefolds for 6D F-theory compactifications indeed fit into the
analogous connected set, and give a good global picture of the set of
allowed Calabi-Yau threefolds, even though additional local non-toric
structure and a variety of tuned elliptic fibration structures are
possible on many toric bases and substantially increase the richness
of the moduli space of 6D theories at a more detailed level.  This
suggests that in 4D as well, the connected set of toric threefold
bases may serve as a good guide to the global structure of the set of
vacua, and is not just a small subset of non-generic special cases.

On each independent run of the one-way Monte Carlo trajectory through
the connected set of toric threefold bases,
we assign a dynamic weight factor to each base in
the blow-up sequence. After
generating a large number of independent random blow-up sequences,
averaging information about the bases across runs using 
the dynamic weight factor can give us
statistical information about the whole set of bases. From this weight
factor, we estimate that the total number of distinct resolvable bases one can
get from blowing up $\mathbb{P}^3$ is approximately $10^{1,964}$,
while the total number of good bases is approximately $10^{253}$.  We
observe, however, that these numbers are likely dramatically
underestimated since the estimated number of bases with larger
$h^{1,1}(B)$ becomes significantly smaller than 1. This underestimate
appears to
occur in large part because our estimations are based on assuming a graph in which
all blow downs return to the starting point, which as we discuss is
generally far from the case.

The dominating class of good bases
that we encounter are the ``end point'' bases, which
are the end points of each sequence of blow ups. They typically have
large $h^{1,1}(B)$, on the order of $10^3$--$10^4$, with a number of
$E_8$, $F_4$, $G_2$ and SU(2) non-Higgsable gauge group
factors. These end
point bases are not random; the generic elliptic CY 4-fold $X$
over them resembles the mirror fourfold of generic elliptic CY 4-folds
over simple bases, in terms of Hodge numbers. The dominance 
of these specific structures
in the
whole set of good bases appears to come
at least in part from the large number of possible flops of these bases, and their
robustness against a sequence of random contractions and blow ups,
though there also appears to be significant multiplicity of bases with
common Hodge numbers and more disparate detailed structure beyond flops.

The set of resolvable bases has fewer clear organizing principles that
we have been able to detect. However, by performing multiple blow-down
operations on various bases reached in the Monte Carlo runs,
we find a lot of resolvable bases with $h^{1,1}(B)\sim 50$--200 that
cannot be contracted to another smooth base, and
which are neither Fano
or fiber bundles. This implies that a complete survey of resolvable toric
bases from blowing up starting points should include these ``exotic
starting points''.
Alternatively, as indicated by Mori theory, it may be necessary to
deal more systematically with singular starting bases.
As mentioned above,
these exotic smooth toric starting points make an accurate estimate of the full
number of bases difficult.

The structure of this paper is as follows: in
Section~\ref{sec:toricF}, we provide the essential mathematical
background on F-theory compactification and toric bases. In
Section~\ref{sec:MC}, the details of random blow ups and the weighting
of bases is explicitly described. We present the results of the random
blow ups from $\mathbb{P}^3$ and other bases in
Section~\ref{sec:result}. We further investigate the global structure
of the set of threefold bases and explore the set of exotic starting
points in Section~\ref{sec:global}. Finally, we include further
discussions and conclusions in Section~\ref{sec:con}.

\section{F-theory on the generic elliptic CY 4-folds over toric bases}
\label{sec:toricF}

\subsection{F-theory in the non-Higgsable phase and the allowed singularities}

A summary of geometric techniques describing F-theory on the generic
elliptic CY 4-fold over a toric base can be found in \cite{MC}. Here we
only restate the essential information.

The elliptic CY 4-fold over the base manifold $B$ with effective anticanonical line bundle $-K_B$ is described by a Weierstrass equation:
\be
y^2=x^3+fx+g,\label{Weierstrass}
\ee
where the Weierstrass polynomials $f$ and $g$ are holomorphic sections
of line bundles $\mc{O}(-4K_B)$, $\mc{O}(-6K_B)$. At some loci
$L\subset B$, the determinant $\Delta=4f^3+27g^2$ vanishes.  The
elliptic fiber over such loci $L$ is singular. If $L$ is a
(codimension-one) divisor, then the 7-branes over this locus may give
rise to a gauge group
factor in the 4D supergravity theory after
compactification.

In this paper, we  assume that $f$ and $g$ are generic sections,
such that the order of vanishing of $\Delta$ over each irreducible
codimension-one locus $L$ is minimal. Under this condition, the list
of gauge groups that can arise from the codimension-one locus $L$ is
limited. We list the possible gauge groups with the associated
order of vanishing
of $(f,g,\Delta)$ in Table~\ref{t:NHG}.

\begin{table}
\begin{center}
\begin{tabular}{|c |c |c |c |c |}
\hline
Kodaira type &
ord ($f$) &
ord ($g$) &
ord ($\Delta$) &
gauge group\\ \hline \hline
$III$ &1 & 2  &3 & SU(2) \\
$IV$ & $\geq 2$ & 2 & 4 & SU(3) or SU(2)\\
$I_0^*$&
$\geq 2$ & 3 & $6$ & SO(8) or SO(7) or $G_2$ \\
$IV^*$& $\geq 3$ & 4 & 8 & $E_6$ or $F_4$\\
$III^*$&3 & 5 & 9 & $E_7$\\
$II^*$& 4 & 5 & 10 & $E_8$ \\
\hline
non-min & 4 & 6 & 12 & - \\
\hline
\end{tabular}
\end{center}
\caption[x]{\footnotesize  Table of
non-Higgsable non-Abelian gauge groups and their Kodaira singular fiber type.
For the Kodaira fibers $IV$, $I_0^*$ and $IV^*$, the gauge group is not uniquely determined by the orders
of vanishing of $f, g$. One needs additional monodromy information in
the Weierstrass polynomials to fix the precise gauge group.
Note that actually only the gauge algebra is determined by the Kodaira
vanishing type; we do not concern ourselves with quotients of the gauge group by
discrete factors in this paper and generally loosely refer only to the
gauge group.
}
\label{t:NHG}
\end{table}

Note that for the fiber types $IV$, $I_0^*$ and $IV^*$, the gauge group is
specified by additional information encoded in the ``monodromy cover
polynomials'' $\mu(\psi)$ \cite{Bershadsky-all, Morrison-sn,
Grassi-Morrison-2}. Suppose that the
divisor is given by a local equation $w=0$, then for the case of type
$IV$,
 \be
 \mu(\psi)=\psi^2-(g/w^2)|_{w=0}=\psi^2-g_2.
 \ee
Here, $g_2$ is the order $w^2$ term in an expansion of $g$ around $L
=\{w = 0\}$.
The gauge group given by a
type $IV$ singular fiber is SU(3) if and only if
$g_2$ is a complete square. The case of type $IV^*$ is similar, where
the monodromy cover polynomial is 
\be
\mu(\psi)=\psi^2-(g/w^4)|_{w=0}=\psi^2-g_4.
\ee
When $g_4$ is a complete square, then the corresponding gauge group is $E_6$, otherwise it is $F_4$. For the case of type $I_0^*$, the monodromy cover polynomial is
\be
\mu(\psi)=\psi^3+(f/w^2)|_{w=0}\psi+(g/w^3)|_{w=0}=\psi^3+f_2\psi+g_3.\label{monocover}
\ee
When $\mu(\psi)$ can be decomposed into three factors:
\be
\mu(\psi)=(\psi+a)(\psi+b)(\psi-a-b),
\ee
the corresponding gauge group is SO(8). Otherwise, if it can be decomposed into two factors:
\be
\mu(\psi)=(\psi+a)(\psi^2-a\psi+b),
\ee
the gauge group is SO(7). If it cannot be decomposed at all, then the gauge group is the lowest rank one: $G_2$.

Non-Higgsable Abelian groups arise from the Mordell-Weil group of the
generic elliptic fibration, and have been shown to not appear on toric
bases \cite{NHU(1)s}. Hence we do not need to include them in this
study.

When non-Abelian gauge groups appear, the Weierstrass form
(\ref{Weierstrass}) is singular. In the conventional F-theory/M-theory
duality, the singular fourfold $X_{sing}$ should be resolved to a
smooth one $X_{sm}$, on which M-theory compactifies. In the M-theory
picture with one lower dimension, this corresponds to giving a
v.e.v. to the scalar in the 3D $\mathcal{N}=2$ vector multiplet and
entering the Coulomb branch. When $(f,g,\Delta)$ vanishes to order
$(4,6,12)$ or higher over a codimension-one locus,  such a
resolution does not exist and  the
 F-theory geometry  does not describe a
supersymmetric vacua. On the other hand, there are three types of 
singularities that are not entirely bad:

$\bullet$ Codimension-two (4,6) loci.

When $(f,g,\Delta)$ vanish to order $(4,6,12)$ or higher over  one or more
codimension-two loci on $B$, we can try to blow up these loci and
lower the degree of vanishing of $(f,g)$ to be less than $(4,6)$. If
this blow-up procedure can be done without introducing a codimension-one
(4,6) locus in the process, then we call this base $B$ a ``resolvable
base''. In fact, the numbers $h^0(\mc{O}(-4K))$ and $h^0(\mc{O}(-6K))$
are unchanged after this resolution process. Suppose that the
codimension-two locus can be described by local equations
$s=t=0$. Then $f$ and $g$ can be written as 
\be 
f=\sum_{p}f_p
s^{f_1(p)}t^{f_2(p)}\ ,\ g=\sum_{p}g_p s^{g_1(p)}t^{g_2(p)} \,,
\ee
where $p$ labels different monomials and $f_i(p),g_i(p)$ are natural
numbers. We know that $\forall p$, $f_1(p)+f_2(p)\geq 4$,
$g_1(p)+g_2(p)\geq 6$. Hence after we blow up $s=t=0$, none of the
polynomials in $f$ and $g$ are removed.

In 6D F-theory, the existence of such a codimension-two (4,6) locus
signifies an SCFT sector coupled to gravity. On the tensor branch
where this locus is blown up, the D3 brane wrapping the exceptional
2-cycle on the base gives a massive string in the 6D F-theory. Going
back to the limit where this exceptional 2-cycle is shrunk to zero
volume, this object will become a tensionless string, giving an
infinite tower of massless states.

In 4D F-theory, the correspondence between such codimension-two (4,6)
loci and the 4D $\mathcal{N}=1$ SCFT is not as clear. However,
analogously there can be a D5 brane wrapping the exceptional 4-cycle
after the blow up. Hence in the limit where this exceptional 4-cycle
is shrunk to zero volume, we will have a tensionless string when we
attempt to directly compactify F-theory over the base with a
codimension-two (4,6) locus. 
The physical consequences of the resulting infinite tower of massless
string modes coupled to gravity is not clearly understood in general
4D situations of this type.  

Note that sometimes we have codimension-two (4,6) singularities with
no gauge group involved, for example:
\be
y^2=x^3+(a_0 s^4+a_1 t^4+\dots)x+(b_0 s^6+b_1 t^6+\dots),
\ee
where the coefficients $a_0,a_1,b_0,b_1, \ldots$ are generic functions of
$u$ and there may be other higher order terms in $s$ and $t$. Since
there is no gauge group on $s=0$ or $t=0$, we cannot enter the Coulomb
branch in the 3D M-theory picture.  However, we are free to blow up
the base locus $s=t=0$ without changing the complex structure moduli
of the Calabi-Yau fourfold. After this blow up, the singular locus
$x=y=s=t$ vanishes and we have a smooth Calabi-Yau fourfold.

$\bullet$ Terminal singularities on the Calabi-Yau fourfold.

Sometimes there are non-resolvable singularities on the CY 4-fold that
cannot be resolved to a smooth Calabi-Yau geometry, but which have nothing
to do with a $(4,6)$ singularity. For example, we can write  a
Weierstrass equation in the following form using local coordinates
$(s,t,u)$: \be y^2=x^3+(a_0 s+a_1 t)x+(b_0 s^2+b_1 st+b_2 t^2), \ee
where the coefficients $a_0,a_1,b_0,b_1,b_2$ are generic functions of
$u$. Then there will be a singularity over the locus $x=y=s=t=0$. We
cannot resolve this singularity without changing the canonical class
of the CY 4-fold. This type of terminal singularity appears
generically in the complex structure moduli space for many bases, but we treat these
singularities as acceptable ones. 
Singularities of this type were encountered in a large fraction of the
toric threefold bases found in our earlier Monte Carlo analysis
\cite{MC}.
In a recent paper on elliptic CY
3-folds \cite{Terminal}, these terminal singularities are shown to 
correspond to a finite number of neutral chiral matter fields. The
impact of these terminal singularities in a CY 4-fold is  not yet
well-understood.

$\bullet$ Codimension-three (4,6) locus

It is also a common feature that $f$ and $g$ vanish to order (4,6)
or higher over a codimension-three locus in the base
\cite{Morrison-codimension-3, MC}. This may lead to a
non-flat fibration in the resolution process if there is only one
gauge group involved, see Table~1 in
\cite{Lawrie-sn,Braun:2013nqa}. For example, if there is an $E_7$
gauge group on $s=0$:
\be
y^2=x^3+(a_0 s^3+a_1 s^4+\dots)+(b_0 s^5+b_1 s^6+\dots),\label{E7model}
\ee 
then there is a codimension-three (4,6) locus at $a_0=b_0=s=0$, which
leads to a non-flat fiber at this point \cite{Lawrie-sn}.

There are also other types of codimension-three (4,6) loci giving
rise to non-resolvable singularities, such as the following SU(2)$\times
G_2$ model:
\be
y^2=x^3+s^2 t^2 x+s^2 t^3 b_0,
\ee
where $b_0$ is a function of $u$. After the crepant resolutions
$(x,y,s;\xi_0)$, $(x,y,t;\eta_0)$ and $(y,\eta_0;\eta_1)$ 
($(x,y,s;\xi_0)$ means blowing up $x=y=s=0$ to get an exceptional
divisor $\xi_0=0$), we get
\be
y^2\eta_1=(x^3\xi_0+s^2 t^2 x\xi_0+s^2 t^3 b_0)\eta_0.\label{res1}
\ee
This is still singular at the codimension-3 locus $y=b_0=(x^2+s^2 t^2)x=\xi_0=0$, which cannot be resolved crepantly since there is no gauge group on the base locus $b_0=0$.

There is no way to resolve the singular Calabi-Yau fourfold by blowing
up the base at a point unless $f$ and $g$ vanish to order (8,12) or
higher. Hence we treat such codimension-three (4,6) loci in an equal
way to the terminal singularities, and we keep the bases with them as
good bases. A clearer classification of codimension-three (4,6) loci
and their physical consequences should be done in further research.

In this paper, we are mainly generating resolvable bases and its small
subset: the ``good bases'' without a codimension-two (4,6) locus.
Terminal singularities are generally accepted, and the good bases we
generated in this paper never 
possess any
codimension-three locus where $f$ and $g$ vanish to (8,12) or
higher. 
(This is an empirical observation from the limited set of runs
described here; we are not aware of any reason that
such loci cannot arise in principle.)
Hence we do not worry about codimension-three loci in the
later sections.

\subsection{Toric bases}

We use the following notation for the data of the (smooth, compact) toric base:

$\bullet$ $v_i(i=1,\dots,h^{1,1}(B)+3)$: 3D vectors denoting the 1D
rays in the fan of the toric base. Note that the total number of these
vectors is always equal to $h^{1,1}(B)+3$.

$\bullet$ $D_i$: The divisor corresponding to the ray $v_i$, which is described by the local equation $z_i=0$.

$\bullet$ $\{\sigma\}$: The set of 3D cones in the fan of the toric base. Because the bases are smooth and compact, each cone $\sigma$ has unit volume, and the total number of 3D cones is always equal to $2h^{1,1}(B)+2$.

For toric threefold bases, the set of Weierstrass monomials is a set of lattice points in $\mathbb{Z}^3$:
\be
\fset=\{u\in\mathbb{Z}^3|\forall v_i,\langle u,v_i\rangle\geq -4\},
\ee
\be
\gset=\{u\in\mathbb{Z}^3|\forall v_i,\langle u,v_i\rangle\geq -6\},
\ee
where $v_i$ are the 1D rays in the fan of the toric base. 

The order of vanishing of $f$ and $g$ on a toric divisor $D_i$ corresponding to a ray $v_i$ is 
\be
\bsp
&\text{ord}_{D_i}(f)=\min(\langle u,v_i\rangle+4)|_{u\in\fset},\\
&\text{ord}_{D_i}(g)=\min(\langle u,v_i\rangle+6)|_{u\in\gset},
\end{split}
\ee
We can also write down the order of vanishing of $f$, $g$ and $\Delta$
on a toric curve $D_i D_j$ corresponding to a 2D cone $v_i v_j$:
\be
\bsp
&\text{ord}_{D_i D_j}(f)=\min(\langle u,v_i\rangle+\langle u,v_j\rangle+8)|_{u\in\fset},\\
&\text{ord}_{D_i D_j}(g)=\min(\langle u,v_i\rangle+\langle u,v_j\rangle+12)|_{u\in\gset},
\end{split}
\ee

Hence the set of Weierstrass monomials of a blown up base is
always a subset of the old one, because a new ray is added into
the fan after the blow up. 
We can blow up a toric curve $D_iD_j$ by adding a new ray $v = v_i +
v_j$, which removes monomials in $\fset, \gset$ with $\langle u, v \rangle <
  4, 6$.  Or we can blow up a toric point $D_iD_jD_k$ by adding a new
  ray $v = v_i + v_j + v_k$ and similarly removing monomials.
As we have discussed earlier, if a curve
$D_i D_j$ with $\text{ord}_{D_i D_j}(f)\geq 4,\text{ord}_{D_i
  D_j}(g)\geq 6$ is blown up, then the sets of Weierstrass monomials
will not change. To check whether the new base is resolvable or not
after a blow-up,
one needs to look at the structure of the lattice polytope
$\gset$. If the origin $(0,0,0)$ does not lie on the boundary of
the lattice polytope $\gset$, then after the resolution process
where all the (4,6) curves are blown up, there will not be a
codimension-one (4,6) locus on any divisor. This follows because if
there exists such a divisor corresponding to the ray $v$, then all the
points $u\in\gset$ satisfying $\langle u,v\rangle<0$ will vanish
  and the origin $(0,0,0)$ lies on the boundary plane $\langle
  u,v\rangle=0$ of $\gset$.

In summary, one only needs to check whether $(0,0,0)$ lies on the
boundary of $\gset$ or not to see whether a base is
non-resolvable or resolvable.

We can estimate the Hodge numbers $h^{1,1}$ and $h^{3,1}$ of the generic elliptic fourfold $X$ over a good base $B$ using the toric data and the monomials in $f$ and $g$. For $h^{1,1}(X)$, we use the Shioda-Tate-Wazir formula:

\be
h^{1,1}(X)=h^{1,1}(B)+\text{rk}(G)+1+N(blp),
\ee
where $G$ is the non-Abelian gauge group on $B$. $N(blp)$ is the
number of additional blow ups to resolve the (non-toric)
codimension-two (4,6) locus on divisors with $E_8$ gauge groups. If this codimension-two (4,6) locus is irreducible, then it contributes 1 to $N(blp)$. If it is reducible, then its contribution to $N(blp)$ is the number of irreducible components.

For the Hodge number $h^{3,1}(X)$, we use an approximate Batyrev type formula \cite{MC}:
\begin{eqnarray}
h^{3,1}(X) &\cong &
\tilde{h}^{3, 1}(X)\label{h31}\\
&= &|\fset|+|\gset|-\sum_{\Theta\in\Delta,\dim\Theta=2}l'(\Theta)-4+\sum_{\Theta_i\in\Delta,\Theta_i^*\in\Delta^*, \dim(\Theta_i)=\dim(\Theta_i^*)=1}l'(\Theta_i)\cdot l'(\Theta_i^*)\,.\nonumber
\end{eqnarray}
Here $\Delta^*$ is the convex hull of $\{v_i\}$ and $\Delta$ is the dual polytope of $\Delta^*$, defined to be
\be
\Delta=\{u\in\mathbb{R}^3|\forall v\in\Delta^*\ ,\ \langle u,v\rangle\geq -1\}.
\ee
The symbol $\Theta$ denotes 2d faces of $\Delta$. $\Theta_i$ and
$\Theta_i^*$ denote the 1d edges of the polytopes $\Delta$ and
$\Delta^*$. $l'(\cdot)$ counts the number of integral interior points
on a face.

\section{A one-way
Monte Carlo approach to blowing up toric base threefolds}
\label{sec:MC}

\subsection{The old approach}

In \cite{MC}, we performed a random walk starting from the base
$\mathbb{P}^3$. In that original algorithm, at each step we do a
random blow up or blow down with equal probability for each choice. We
never allow a base with codimension-one or codimension-two (4,6) locus
in the process. This effectively excludes the bases with
non-Higgsable $E_8$ factors,
because they can only be achieved by passing through bases with
codimension-two (4,6) loci on the divisors containing $E_8$. The
information such as Hodge numbers ($h^{1,1}$ of the base, $h^{1,1}$
and $h^{3,1}$ of the generic elliptic fourfold over the base) and
gauge groups were tracked for each base over a number of independent
runs of this algorithm. After about 10,000 steps in each run, it turns
out that we entered a region with Hodge numbers in a certain range
and characteristic gauge groups. In total, we gathered the information
of 100,000 bases in this random walk for each run. We ran this process
 100 different times and averaged the data among the runs, providing a
 general distribution of data for Hodge numbers, gauge groups and
possible matter curves. We estimated the total number of bases to be
around $10^{48}$, using the ``bounded random walk'' approach. For
example, if we want to know the number of bases with $h^{1,1}(B)=7$,
we can do a random walk with limit $h^{1,1}(B)\leq 7$. We compute the
ratio of the number of bases with $h^{1,1}(B)=7$ to the number of
bases with $h^{1,1}(B)=2$: $\mathcal{N}(7)/\mathcal{N}(2)$. Then
because we know that the number of bases with $h^{1,1}(B)= 2$ that is
connected to $\mathbb{P}^3$ is 27, we can estimate the number of bases
with $h^{1,1}(B)$ to be
$27\cdot\mathcal{N}(7)/\mathcal{N}(2)$. Similarly we can repeat the
process with higher and higher upper bound on $h^{1,1}(B)$.
(This estimation needs to be done in this incremental bounded fashion,
since the unbounded random walk rapidly moves to larger $h^{1, 1}$
and never returns to the smaller values for which the explicit
enumeration is known.)

There are two drawbacks of this approach. First, the set of bases we
generated is incomplete, since a lot of allowed bases can only be generated by
going through bases with codimension-two (4,6) loci and are not
reached by this algorithm. As a 2D example,
the base that leads to the self-mirror elliptic Calabi-Yau threefold
with $h^{1,1}=h^{2,1}=251$ can only be generated by blowing up the
Hirzebruch surface $\mathbb{F}_{12}$ through these 
kind of codimension-two (4, 6)
regions
\cite{mt-toric,Hodge}. Specifically, we expect that a lot of F-theory
bases will give rise to non-Higgsable $E_8$ gauge groups, and these
are not included in the old Monte Carlo runs. Second, the graph with
these $10^{48}$ bases has low connectivity, so that each run appears
to ``thermalize'' in a local region of the landscape.  For example, it
happened that in some particular runs, most of the 100,000 gauge groups
have some special characteristic gauge group such as SO(8) or $E_7$
that cannot be seen from most other runs. This fact indicates that the
random walks do not necessarily capture the global average features of
the whole graph, and that the 100,000 steps in each run may be an
inefficient way to estimate global averages over the full landscape.

\begin{figure}
\centering
\includegraphics[height=6cm]{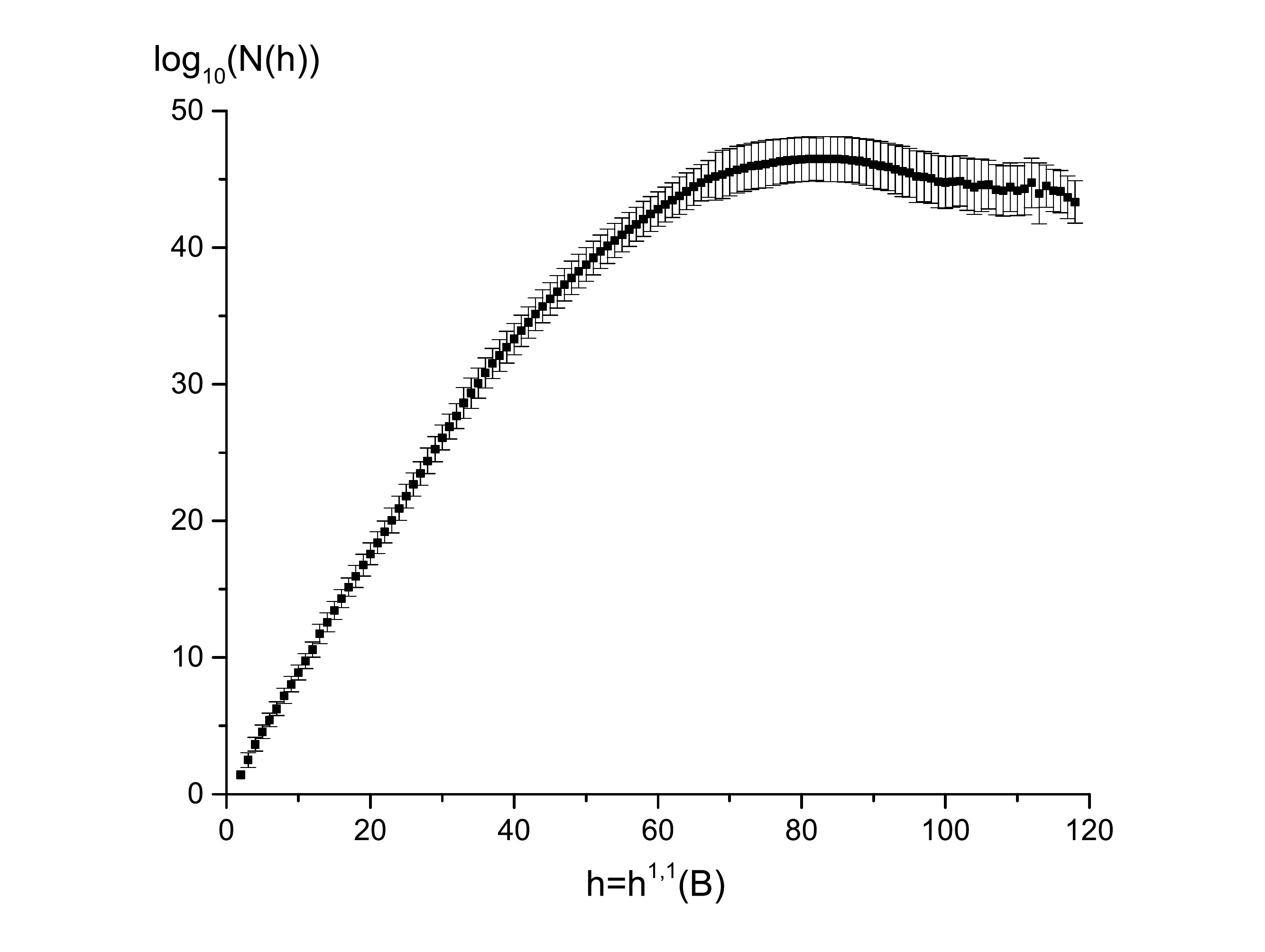}
\caption[E]{\footnotesize $\log_{10}N$ of the number of bases with certain $h^{1,1}(B)$ in the old Monte Carlo approach. The average and error bar are computed among 100 runs.} \label{logNold}
\end{figure}

Similar to the 2D toric approach, we want to cover as many as good
bases as we can.  As discussed above, ``good'' here means that the
generic fibration over the base has no (4,6) locus on toric
curves. There may be (4,6) loci on (non-toric) curves on a divisor
with $E_8$, similar to the cases with $(-9/-10/-11)$-curves on the 2D
toric bases. They are allowed since we can get a base without these
(4,6) loci by simply blowing up these loci. The gauge group on the
divisor remains $E_8$, which is not affected by the process. Hence we
want to run a Monte Carlo program that is allowed to pass through
these threefold bases with toric codimension-two (4,6) loci that are
resolvable into good bases. Then we wish to estimate the number of
these resolvable bases and the number of good bases, and to determine
the generic features of bases in this ensemble. The old random walk
approach will not work, because the number of good bases is negligible
compared to the vast number of resolvable bases, so the possibility of
finding a good base in a random walk is almost zero.

\subsection{The new random blow up algorithm}

\subsubsection{The algorithm}

In this paper, we try a different Monte Carlo approach. We start from
a base, say $a_1=\mathbb{P}^3$, and then randomly generate a blow-up
sequence from it, giving bases $a_2, a_3, \ldots$. 
After $(k-1)$ blow ups, $h^{1, 1}(B) =h^{1, 1} (a_1)+k-1$; so for $a_1
=\P^3,h^{1, 1} (a_k) = k$.
Resolvable codimension-two singularities are allowed
throughout the process, including (4,6) singularities on toric curves
and on curves on $E_8$ divisors. The numbers of possible ways of blowing
up and down from a base $a_i$ are recorded, and are called
$N_{\rm out}(a_i)$ and $N_{\rm in}(a_i)$. These numbers are relatively easy to
compute.  The number of possible blow-downs $N_{\rm in}$ is easier to compute than
the number of blow ups, because from the definition of resolvable
bases, any smooth blow down of a resolvable base gives another resolvable
base. 
The number of blow-ups $N_{\rm out}(a_i)$ can be evaluated by checking
for each possible blow
up of toric curves and points 
whether the point (0, 0, 0) is on the boundary of the set of monomials
in $\gset$, as discussed above.
In the counting of $N_{\rm out}(a_i)$ and $N_{\rm in}(a_i)$, we only
count the number of toric bases up to an $SL(3,\mathbb{Z})$
transformation on the toric fan. Note that in the older work
\cite{MC}, we introduced a ``symmetry factor'' to correctly count the
non-isomorphic bases (see Section~2.6). But here we explicitly
construct the bases after each possible blow-up/down and compare
them\footnote{In the actual program, we only check the isomorphism
  among the bases after blow up/downs for the base with $h^{1,1}<10$,
    since an isomorphism between the resulting bases is nearly
    impossible to occur for a general base with $h^{1,1}\geq
    10$. This fact was already verifed in \cite{MC}, where we checked
    that the symmetry factor for a base with $h^{1,1}\geq 10$ is
    essentially always
    equal to one.
In principle, a symmetry factor can also be used in the current
approach, which would give equivalent results, but would similarly
become irrelevant as soon as $h^{1,1}$ gets larger than 10 or so.}.

At each step in the blow-up process, we pick one of the $N_{\rm out}$
possible
blow ups, choosing each with equal probability $1/N_{\rm out}$.  After
some finite number of blow ups, this sequence terminates at an ``end point'' where
there is no possible blow up to another resolvable base. This end
point has to be good from definition, and not just resolvable,
because there cannot be any
toric (4,6) curve to be blown up; otherwise we can always blow up this
toric (4,6) curve to get a base with exactly the same sets
$\fset,\gset$. As discussed in the introduction, we know that this
procedure must in principle terminate as the number of toric threefold
bases is finite; in practice it generally terminates after somewhere
between 3000 and 13000 blow ups.

\subsubsection{Weighting factors}

To correctly take into account the unequal possibilities of entering
each branch, we introduce a dynamic weight for each node $a_n$ from
the path $p=a_1\rightarrow a_2\rightarrow\dots\rightarrow a_n$:
\be
D(p=a_1\rightarrow a_n)=\prod_{i=1}^{n-1}\frac{N_{\rm out}(a_i)}{N_{\rm in}(a_{i+1})}.\label{dweight}
\ee
We call the subscript $i$ of node $a_i$ the ``layer number'' of
$a_i$. We claim that (\ref{dweight}) gives the correct weight of each
node such that the weighted possibility of getting each node $a_n$ sums
up to 1, under the assumption that the whole graph can be scanned by moving up
from one initial node $a_1$.

We prove this by induction. Assume that this holds for all the nodes with layer number less than or equal to $k-1$, so that
\be
\sum_{p\rightarrow a_{k-1}}D(p\rightarrow a_{k-1})\mathcal{P}(p\rightarrow a_{k-1})=1\label{dynweight}
\ee
for all the nodes $a_{k-1}$ with layer number $k-1$. Here the sum is over all the paths leading to the node $a_{k-1}$, and $\mathcal{P}(p\rightarrow a_{k-1})$ is the probability of this path.

Now suppose that a node $a_k$ with layer number $k$ has $m_k$ nodes $a_{k-1,1},a_{k-1,2},\dots,a_{k-1,m_k}$ linked to it. Then the sum
\be
\bsp
&\sum_{p\rightarrow a_{k}}D(p\rightarrow a_{k})\mathcal{P}(p\rightarrow a_{k})
\\=&
\sum_{q=1}^{m_k}\sum_{p\rightarrow a_{k-1,q}}D(p\rightarrow a_{k-1,q})\cdot\frac{N_{\rm out}(a_{k-1,q})}{N_{\rm in}(a_k)}\cdot\mathcal{P}(p\rightarrow a_{k})\\
=&\sum_{q=1}^{m_k}\sum_{p\rightarrow a_{k-1,q}}D(p\rightarrow a_{k-1,q})\cdot\frac{N_{\rm out}(a_{k-1,q})}{m_k}\cdot\mathcal{P}(p\rightarrow a_{k-1,q})\cdot\frac{1}{N_{\rm out}(a_{k-1,q})}\\
=&\sum_{p\rightarrow a_{k-1,q}}D(p\rightarrow a_{k-1,q})\mathcal{P}(p\rightarrow a_{k-1,q})\\
=&1.
\end{split}
\ee
Here we used the fact that $D(p\rightarrow a_{k})=D(p\rightarrow a_{k-1,q})\cdot\frac{N_{\rm out}(a_{k-1,q})}{N_{\rm in}(a_k)}$ and $\mathcal{P}(p\rightarrow a_{k})=\mathcal{P}(p\rightarrow a_{k-1,q})\cdot\frac{1}{N_{\rm out}(a_{k-1,q})}$.

The identity (\ref{dynweight}) simply holds for $k-1=1$, which
completes the proof.

\vspace{0.5cm}

From the dynamical weight factor, we can estimate the average of
quantities across the graph.  For a given property $f (a_k)$ of a node
in the graph, such as the number
of outgoing edges, we can determine the total of $f$ across all nodes
at level $k$ as
\begin{equation}
\sum_{a_k}f (a_k)    = \langle f (a_k)D (k) \rangle
= \sum_{{\rm paths} \;p = a_1 \rightarrow a_n} 
f (a_k) D (p \rightarrow a_k){\cal P} (p \rightarrow a_k) \,,
\end{equation}
which can be estimated by simply averaging $f (a_k)D (k)$ across a
large number of one-way Monte Carlo runs from the initial node $a_1$
to level $k$.

In particular,
we can directly estimate the number of nodes at level $k$ in the graph as
\be
N_{\rm nodes}(k)=\langle 1 \cdot D (k) \rangle \approx
\frac{1}{N(p)}\sum_{i=1}^{N(p)}D(p\rightarrow k),\label{ND}
\ee
where $N(p)$ is the total number of sampling branches and $D(p\rightarrow k)$ is the weight factor when a branch $p$ reaches the layer $k$. If a branch never reaches layer $k$ then we take $D(p\rightarrow k)=0$.

We can estimate the average of a quantity $f (k)$ across all nodes
$a_k$ at level $k$ by dividing the
total by the number of nodes
\begin{equation}
\langle f (k) \rangle_D \equiv \frac{\langle f (a_k)D (k) \rangle}{
  \langle D (k) \rangle} \,.
\end{equation}

This gives an alternative expression relating
the total number of nodes in layer $k$ to the total number of nodes in layer $k-1$,
\be N_{\rm nodes}(k)=\frac{\langle N_{\rm out}(k-1)\rangle_D}{\langle
  N_{\rm in}(k)\rangle_D}\cdot N_{\rm nodes}(k-1)  \,.\label{Nnodes} \ee
Following through the definitions shows that this estimate is
precisely equivalent to (\ref{ND}), even for a finite number of
samples $N (p)$.

Finally, we can compute the number of good bases $N_{\rm good}(k)$ out of
the resolvable bases, simply by multiplying the relative weight factor
on $N_{\rm nodes}(k)$:
\be
N_{\rm good}(k)=N_{\rm nodes}(k)\times\frac{\sum_{a_k\textrm{ is good}}D(p\rightarrow a_{k})}{\sum_{a_k} D(p\rightarrow a_{k})}.\label{Ngood}
\ee
To estimate this quantity with a finite set of runs, we can simply use
(\ref{ND}), where  trajectories that do not
reach a good base at level $k$ contribute 0,  i.e. simply averaging the sampled
value of $D (k)$ over
the good bases at that level and multiplying by the fraction of
trajectories that reach a good base at level $k$.

\subsubsection{Systematic issues with estimating the number of bases}
\label{sec:systematic-issues}

\begin{figure}
\centering
\includegraphics[height=4cm]{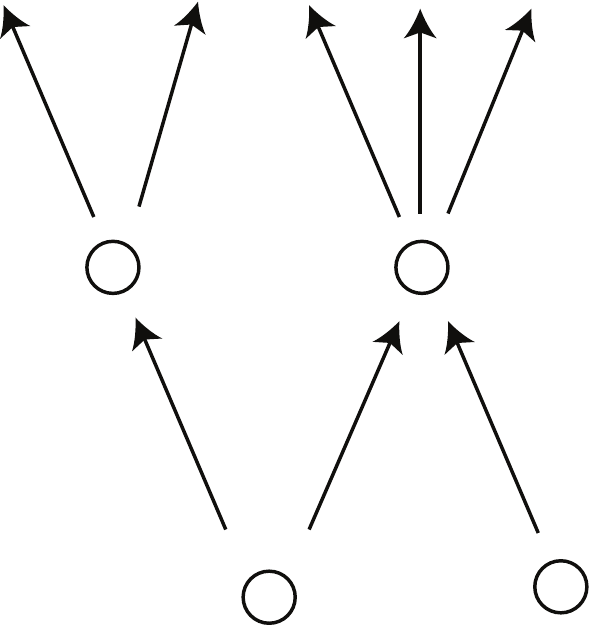}
\caption[E]{\footnotesize A typical graph where there are side
  branches entering the tree from the starting point. The $N_{\rm in}$ of
  the right node in the second layer should be 1 instead of 2, if we
  are interested in computing averages across the tree of nodes
  accessible from only the left starting point.} \label{f:graph1}
\end{figure}

There are several reasons that the methodology described so far leads
to a systematic underestimate of the number of bases.  One key issue
is that we have assumed in the analysis above that for each base $a_k$
that is reached from a sequence of blow ups from the starting point
$a_1$, every acceptable blow-down of $a_k$ can also be reached by a
sequence of blow ups from $a_1$.  A problem arises, however, when the
graph is like the one shown in Figure~\ref{f:graph1}. If there are
side branches entering the tree, the estimated number of nodes will be
lower than the correct one, since the measured $\langle N_{\rm
  in}(k)\rangle$ will be higher than its correct value
when considering only blow-ups of $a_1$.

Because we are counting the number of bases one can get from blowing
up the starting point, one should only count the $N_{\rm in}(a_k)$ of
a node for the blown down bases $b_{k-1}$ of $a_k$ that can be
contracted to the starting point by a sequence of blow downs. This new
$N'_{\rm in}(a_k)$ is always smaller or equal than $N_{\rm in}(a_k)$.
The discrepancy between $N'_{\rm in}(a_k)$
and $N_{\rm in}(a_k)$ depends upon both the number of possible other
starting point bases (bases that do not admit a smooth blow down) that
connect to the graph after blowing up, and also upon the degree of
connectivity of the graph.
If the total number of
possible starting points is order one, or much smaller than the total
number of bases --- as it is in six dimensions, where only the
Hirzebruch surfaces and $\P^2$ are minimal starting points --- then
the estimation of the total number of bases in the ensemble should be
reasonably accurate for one or more of the starting point bases.  If,
however, the number of minimally contracted bases is very large, this
can lead to a dramatic underestimate of the total number of bases
using this algorithm.   Even if the number of possible starting bases
is large, however, it only causes a significant problem with counting
if a large fraction of the incoming edges to the nodes $a_k$ reached
from blowing up $a_1$ lead only to other starting points besides $a_1$.

Focusing on the latter issue of connectivity,
we can try to make a simple
estimate of $N'_{\rm in}(a_k)$ by
checking for each possible blow down whether the contracted ray is one of the rays on the starting
point base or not. 
If it is not one of them, then we can
add it into an estimated value $\tilde{N}'_{\rm in}(a_k)$, but otherwise we do not. The motivation  for
dropping these contractable rays
is that if we contract one of the
rays on the starting point, then in general the base may
no longer contain
a set of rays that are linearly equivalent to the set of rays on the
starting point by an $SL(3,\mathbb{Z})$ transformation. It is
possible, however, that after we contract a ray that
corresponds to a ray on
the starting point, there may still exist another configuration of the
starting base somewhere else to which the base may be contracted. Hence this estimated $\tilde{N}'_{\rm in}(a_k)$ may
be smaller than the actual value
 $N'_{\rm in}(a_k)$.
On the other hand, there can also be bases that still include the rays
of the starting point base $a_1$ but cannot be contracted to $a_1$ for
other reasons; these will cause $\tilde{N}'_{\rm in}$ to overestimate
 $N'_{\rm in}$.
As we see in the next section, the distinction between $N_{\rm in}$
and $\tilde{N}'_{\rm in}$
amounts to a fairly minor  difference in numerical results in the
one-way Monte Carlo computations, so this crude estimation does not
detect any significant under or overcounting due to a misestimation of $N'_{\rm
  in}$. On the other hand, as we discuss in more detail in
Section~\ref{sec:global}, it seems that there are many other starting
point bases possible at large $h^{1, 1}(B)$ that are ``dead ends''
reached when we blow down along random incoming edges of the
$N_{\rm in}$ possibilities for $a_k$, so that both $N_{\rm in}$ and
$\tilde{N}'_{\rm in}$ are likely substantially over estimating the
correct value $N'_{\rm in}$ that would need to be used to correctly
determine the number of nodes in the graph.

\begin{figure}
\centering
\includegraphics[height=6cm]{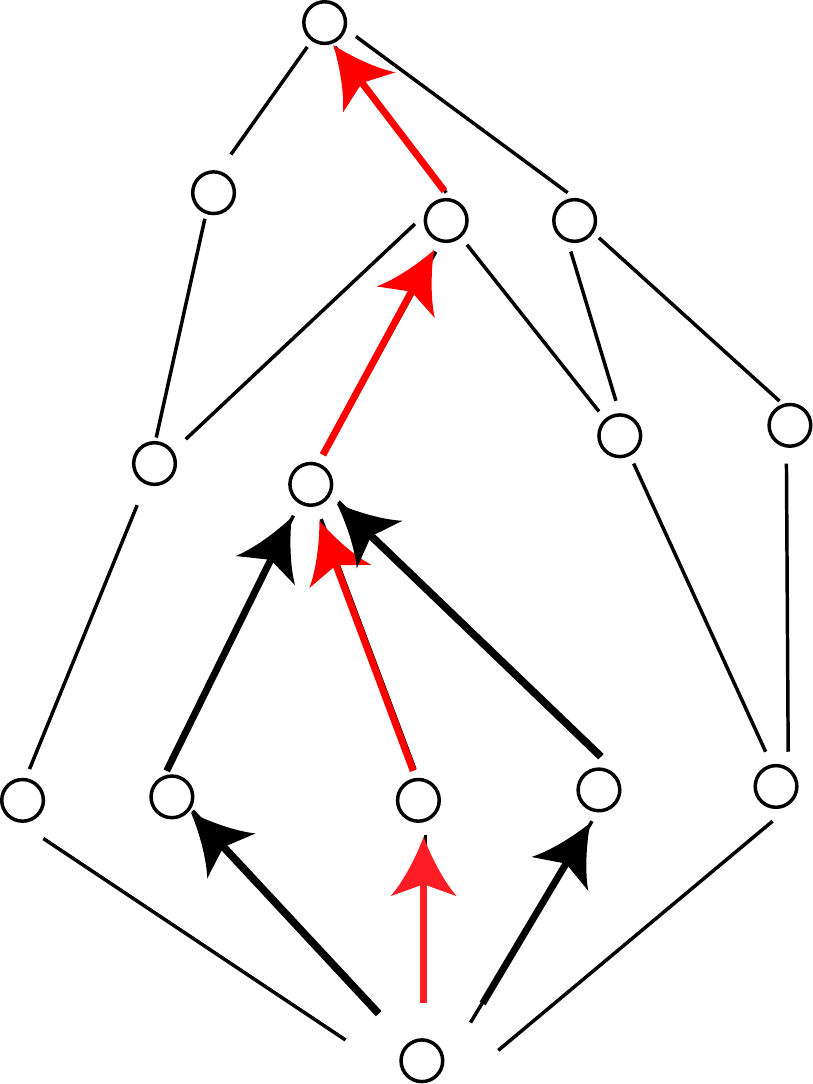}
\caption[E]{\footnotesize An example of a common branch in a random
  graph. Only from the information of $N_{\rm out}$ and $N_{\rm in}$ along
  this branch, we get a substantially underestimated total number of
  nodes in the top layer: $\frac{5}{27}$} \label{f:strange}
\end{figure}

Finally, there is one further issue in this one-way Monte Carlo
approach that results in a smaller estimation of the total number of
bases, even if there are no additional starting points possible or
additional edges entering the tree.  While in principle the estimate
(\ref{ND}) gives an accurate estimate of the number of nodes when
carried out over many runs of the one-way Monte Carlo, this estimate
may only be accurate when enough runs are done to completely explore
the set of possibilities, which is practically impossible as the
number of trajectories through the graph grows exponentially in $k$.
In practice, the most probable branches of the blow-up tree that we
enter are the ones with small weight factors, which lead to a small
estimated number of $N_{\rm nodes}$ and $N_{\rm good}$.  As an
example, consider the red branch shown in Figure~\ref{f:strange}. If
we only do one random blow up sequence through this graph, then we
have a 60\% possibility to enter this branch or the other two branches
besides it. Applying (\ref{Nnodes}) repeatedly along this path, we get
that the estimation of the number of nodes in the top layer is given by
the weight factor
$D =5/27$ rather than 1.
Thus, most of the time a random blow-up algorithm on this graph would
give an expected number of nodes of $< 0.2$ at the top level.  This is
  compensated by low-probability paths with large weight; for example,
  the path along the left side of the graph has probability $0.1$ but
  gives a weight factor $D = 10/3$.  While indeed one can check that
  the expectation value over all paths in this graph is indeed
$\langle D \rangle = 1$, in a larger graph, such as one composed of
  many iterated copies of this graph, the distribution of $D$ values
  becomes highly asymmetric, and typical paths will give much lower
  values of $D$ than the idealized average.  For graphs with a simple
  regular structure, such as an iterated version of the graph in
Figure~\ref{f:strange},
this systematic underestimate can be compensated for by observing that
the distribution of $D$'s takes a lognormal form.  An example of how
this can be corrected for in the case of a simple toy model is worked
out in detail in
Appendix A.  Because the graph of connected threefold toric bases that
we are exploring does not have a simple regular structure, it is
unclear to what extent we can systematically compensate for this
effect, but as we discuss in the following section this may be
possible at least for local perturbations around a given trajectory.

To summarize the situation, there are several reasons why the
approximation methods described in the previous subsection can give a
systematic underestimate of the total number of nodes that can be
reached by sequential blow-ups of a given toric threefold base $a_1$.
We have not identified a clear and simple way of accurately
compensating for the systematic underestimates just described, so the
results presented here should be taken as lower bounds, with a more
accurate estimation requiring an improved methodology for dealing with
incoming branches such as in Figure~\ref{f:graph1} or the effect of
more likely branches as in Figure~\ref{f:strange}. We will discuss
this issue further in Section~\ref{sec:error} in the context of the
experimental data.

\section{Results}
\label{sec:result}

\subsection{Blowing up $\mathbb{P}^3$}

\begin{figure}
\centering
\includegraphics[height=6cm]{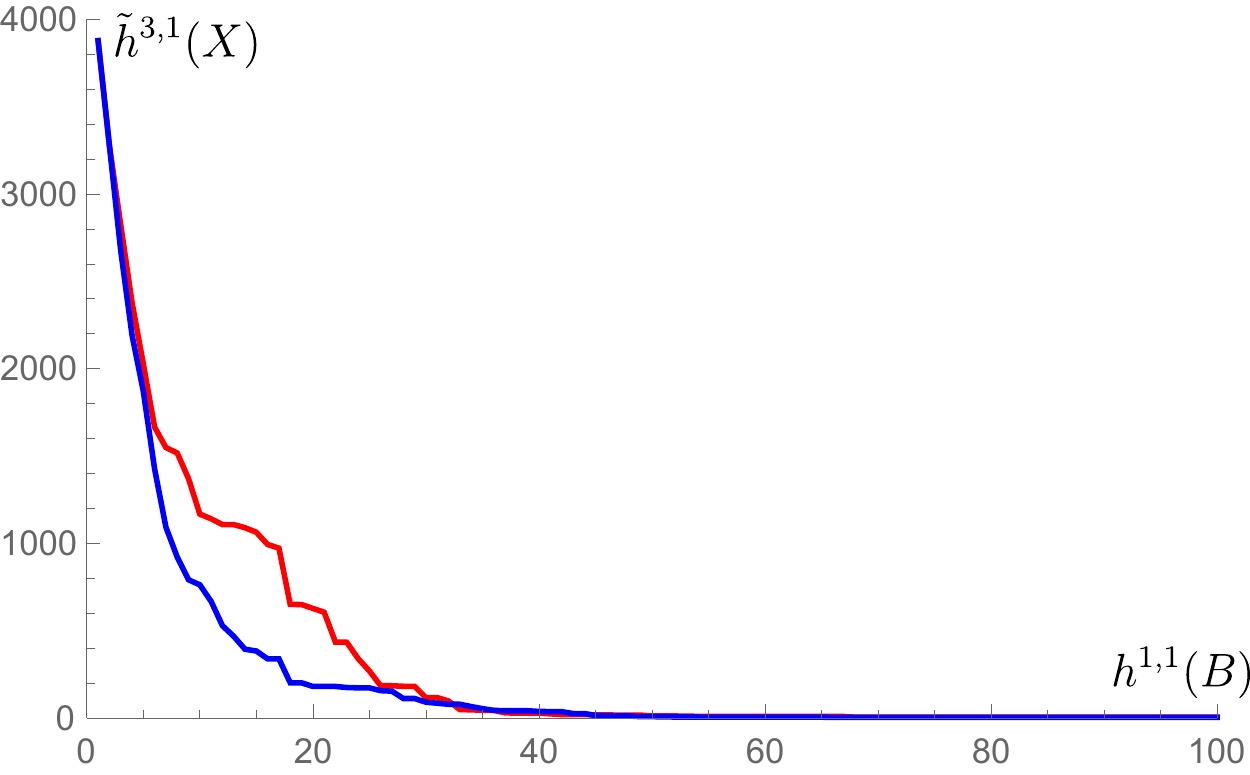}
\caption[E]{\footnotesize The change in $\tilde{h}^{3,1}(X)$ as a function of $h^{1,1}(B)$ for two random blow up sequences. The number of Weierstrass moduli drops quickly.} \label{f:h11Bh31}
\end{figure}

We have done 2,000 random blow up sequences starting from
$\mathbb{P}^3$. We plot $\tilde{h}^{3,1}(X)$ of a generic elliptic
fourfold over the base $B$ (see (\ref{h31})) for some example blow up sequences as a function of $h^{1,1}(B)$ in Figure~\ref{f:h11Bh31}. As we can see, the number of
 Weierstrass moduli quickly drops to a very small number after
  about 50 blow ups.
  
The sequences terminate after somewhere between 3,000 and 13,000
individual blow ups.  We summarize the results of this numerical
experiment, beginning with the set of good bases encountered and then
addressing the estimation of the number of nodes in the full graph
using the methods described in the previous section.

\subsubsection{Good bases and end points}

In the first 20 steps of the blow-up trajectory, many of the sequences
encounter ``good'' bases without codimension-two (4, 6) loci.  After
this, all the sequences developed codimension-two (4, 6) toric curves,
which continue to dominate the base geometry until the process
converges at a terminal ``end point'' base that cannot be blown up
further, which as discussed above is always ``good''.  An interesting
feature of the results of this computation is that the Hodge numbers
of the good end point bases appear to be clustered at certain specific
isolated values, with some geometric significance.

We list the unweighted number of good bases $n_{\rm good}$ encountered
among the 2,000 runs at each level $k$, with $h^{1, 1}(B) = k$, in
Table~\ref{t:distribution}.  This gives some estimation of the
probability of finding a specific end point base from a random
sequence.  For those values with $h^{1,1}(B)<20$, the good bases
  encountered arise near the beginning of the blow-up sequence, and
  these are never terminal end point bases.  The number of good bases
  decreases when $h^{1,1}(B)$ increases, since codimension-two (4,6)
  loci appear during the blow up process. The remaining good bases
  arise as end points.  These have very large $h^{1,1}(B)$ and are
  concentrated at sporadic values of $h^{1,1}(B)$.

\begin{table}
\centering
\begin{tabular} {|c|c||c|c||c|c||c|c|}
\hline
$h^{1,1}(B)$ & $n_{\rm good}$ & $h^{1,1}(B)$ & $n_{\rm good}$ & $h^{1,1}(B)$ & $n_{\rm good}$ & $h^{1,1}(B)$ & $n_{\rm good}$ \\
\hline
1 & 2000 & 2 & 2000 & 3 & 2000 & 4 & 1900\\
5 & 1645 & 6 & 1259 & 7 & 846 & 8 & 545\\
9 & 283 & 10 & 129 & 11 & 48 & 12 & 26\\
13 & 13 & 14 & 8 & 15 & 4 & 16 & 1\\
\hline
1317 & 1 & 1727 & 17 & 1799 & 4 & 1882 & 1\\
1943 & 198 & 2015 & 41 & 2047 & 23 & 2057 & 139\\
2186 & 44 & 2199 & 10 & 2249 & 315 & 2303 & 306\\
2395 & 10 & 2399 & 31 & 2491 & 6 & 2591 & 205\\
2599 & 17 & 2623 & 64 & 2636 & 6 & 2661 & 29\\
2821 & 4 & 2824 & 16 & 2891 & 1 & 2915 & 2\\
2943 & 1 & 2961 & 21 & 2999 & 40 & 3037 & 9\\
3071 & 3 & 3086 & 112 & 3157 & 1 & 3247 & 2\\
3276 & 4 & 3295 & 2 & 3374 & 4 & 3401 & 4\\
3422 & 34 & 3498 & 3 & 3539 & 2 & 3599 & 2\\
3658 & 12 & 3686 & 55 & 3739 & 1 & 3741 & 2\\
3789 & 1 & 3811 & 3 & 3817 & 1 & 3887 & 27\\
3992 & 1 & 4049 & 4 & 4211 & 1 & 4274 & 1\\
4373 & 25 & 4375 & 3 & 4394 & 21 & 4468 & 4\\
4520 & 1 & 4741 & 1 & 4748 & 1 & 4913 & 5\\
4939 & 10 & 4946 & 1 & 5143 & 21 & 5356 & 1\\
5383 & 5 & 5503 & 2 & 5522 & 1 & 5623 & 1\\
5878 & 1 & 5989 & 6 & 6143 & 2 & 6440 & 2\\
6784 & 1 & 6802 & 5 & 6911 & 8 & 6945 & 1\\
7373 & 1 & 7498 & 2 & 7526 & 1 & 7909 & 7\\
8111 & 3 & 8230 & 1 & 8435 & 1 & 8938 & 1\\
8980 & 5 & 8999 & 1 & 10124 & 3 & 11341 & 2\\
12631 & 1 & - & - & - & - & - & -\\
\hline
\end{tabular}
\caption[E]{\footnotesize The number of good bases
encountered for each $h^{1,1}(B)$ among the 2,000 runs, without
counting the weight factor.
Bases at small $h^{1, 1}(B) < 20$ are not end points, all bases with
large $h^{1, 1}(B) > 1000$ are end points of the algorithm.
} \label{t:distribution}
\end{table}

One can see from Table~\ref{t:distribution} that
 for bases with $h^{1, 1}(B) \leq 3$, all bases are good.  Indeed an
 explicit analysis by hand shows that only blowing up two points or
 curves is insufficient to generate a codimension-two (4, 6)
 locus.\footnote{Note that codimension-two (4, 6) loci arise more
   easily on base threefolds than on base surfaces.  For base
   surfaces, the vanishing order of $f, g$ at the intersection of two
   curves is simply the sum of the vanishing orders on the curves, so
   a (4, 6) point on a base surface only arises at the intersection of
 two curves involving NHCs; for base threefolds on the other hand,
 such as some examples with $h^{1, 1}(B) = 4$, it is possible to have
 $(f, g)$ vanishing to orders (4, 6) on a toric curve even when $f, g$
 do not vanish to any order on the divisors that intersect on the
 curve.}
As $h^{1,1}(B)$
increases above $k = 3$,
  the number of good bases among the 2,000 runs decreases quickly, because
there are an increasing number of ways to generate
bases with codimension-two (4,6)
  loci after a number of blow ups.

The other set of good bases encountered in the runs is the set of
scattered end points with $h^{1,1}(B)>1,000$, whose number indeed adds
up to 2,000, as one can check explicitly.  While some $h^{1, 1}(B)$
only arise as the end points of one or a few runs, there are sharp
spikes in the distribution associated with particular values that
arise frequently; of the 2,000 runs, the number of distinct values of
$h^{1, 1}(B)$ at the endpoint is only 89.  It is interesting to study
the structures of the bases at the most prominent spikes, for example
$h^{1,1}(B)=1943$, 2249, 2303 and 2591, each of which arises for
roughly 10\% or more of the blow-up trajectories. It turns out that
for all the end point bases we found with each  specific value of
 $h^{1,1}(B)$, the non-Higgsable gauge
group contents are the same. For $h^{1,1}(B)=1943$, 2249, 2303 and
2591, the gauge groups are
$G=E_8^{29}\times F_4^{81}\times G_2^{216}\times SU(2)^{324}$,
$E_8^{33}\times F_4^{94}\times G_2^{250}\times SU(2)^{375}$,
$E_8^{34}\times F_4^{96}\times G_2^{256}\times SU(2)^{384}$ and
$E_8^{38}\times F_4^{108}\times G_2^{288}\times SU(2)^{432}$
respectively. 
This general structure is a common feature of the end point bases, since
$f$ almost always vanishes to degree 4 on every divisor
(i.e. $\mc{F}=\{(0, 0, 0)\}$), so that the
most common non-Higgsable gauge group
factors will be SU(2), $G_2$, $F_4$ and
$E_8$, corresponding to the cases where $g$ vanishes to degree 2,3,4
and 5 respectively.  
(This same structure arises for elliptic Calabi-Yau threefolds over
base surfaces, where at large $h^{1, 1}(B_2)$ the geometry is
dominated by ``$E_8$ chains'' containing the non-Higgsable clusters
$-12, -5$, and $(-3, -2)$, which carry the gauge groups $E_8, F_4$,
and $G_2 \times SU(2)$ respectively
\cite{mt-toric,Hodge}.)
The gauge groups
SU(3) and SO(8) also appear
in some places, but infrequently;
for example the bases with $h^{1,1}=2999$ have gauge groups
$E_8^{44}\times F_4^{125}\times G_2^{332}\times SU(3)\times
SU(2)^{500}$. 
An empirical formula for the number of gauge group
factors SU(2), $G_2$,
$F_4$ and $E_8$  in terms of $h^{1,1}(B)$ 
goes roughly as 
\be
N_{SU(2)}\cong\left[\frac{h^{1,1}(B)+1}{6}\right]\ ,\ N_{G_2}\cong\left[\frac{h^{1,1}(B)+1}{9}\right]\ ,\ N_{F_4}\cong\left[\frac{h^{1,1}(B)+1}{24}\right]\ ,\ N_{E_8}\cong\left[\frac{h^{1,1}(B)}{68}\right].\label{Ngaugegroups}
\ee

While the non-Higgsable gauge groups appear to be quite uniform across
end points with common $h^{1, 1}(B)$, the non-Higgsable cluster
structures on the bases with the same $h^{1,1}(B)$ are very different;
this can be easily checked by looking at the different total number of
non-Higgsable clusters and their different sizes. These bases also
can
have different convex hulls of the fan, hence they are not always
related by
a series of flops. 

A potentially
very interesting finding is that the Hodge numbers of the elliptic
Calabi-Yau fourfolds associated with end point bases at large
$h^{1,1}(X)$ seem to give the mirror Hodge numbers to those
of various generic elliptically fibered Calabi-Yau fourfolds over bases
with small $h^{1,1}(B)$. 
For example, some of the bases with
$h^{1,1}(B)=2303$ give $h^{1,1}(X)=3878$ and $h^{3,1}(X)=2$. Since
a generic elliptic fibration over $\mathbb{P}^3$ gives an $X$ with
$h^{1,1}(X)=2$ and $h^{3,1}(X)=3878$, these look exactly like 
mirror Calabi-Yau fourfold pairs\footnote{The Hodge numbers of generic
  elliptic Calabi-Yau fourfolds are also computed in
  \cite{Mohri}.}. There are also other bases with $h^{1,1}(B)=2303$ that
give $h^{1,1}(X)=3877$ and $h^{3,1}(X)=4$. They are also included in
the dataset of Calabi-Yau fourfolds constructed as hypersurfaces in
weighted projective space using reflexive polytopes
\cite{Kreuzer-Skarke-4d}. 
For the bases with
$h^{1,1}(B)=1943$, the generic elliptic Calabi-Yau fourfold has
 $h^{1,1}(X)=3277$ and $h^{3,1}(X)=3$,
which exactly looks like the dual of a generic elliptic fibration over
the generalized
Hirzebruch threefold $\tilde{\mathbb{F}}_0$, which is
$\mathbb{P}_1\times\mathbb{P}_2$. We list many of these interesting cases
in Table~\ref{t:endpoints}.  In this table, the
generalized Hirzebruch threefold
$\tilde{\mathbb{F}}_n$ is a $\mathbb{P}_1$ bundle over $\mathbb{P}_2$
with total twist $n$. In toric language, the 1D rays in the fan are
$v_1=(1,0,0), v_2=(0,1,0), v_3=(0,0,1), v_4=(0,0,-1)$ and
$v_5=(-1,-1,-n)$. The 3D cones in the fan are $\{v_1 v_2 v_3, v_1 v_5
v_3, v_2 v_5 v_3, v_1 v_2 v_4, v_1 v_5 v_4, v_2 v_5 v_4\}$.
Note that in other cases the end point bases support elliptic
Calabi-Yau fourfolds that seem to be dual to other Calabi-Yau
fourfolds with small $h^{1, 1}(X)$ but that are not elliptically
fibered.  For example,
 the bases with $h^{1,1}(B)=2249$
give Calabi-Yau fourfolds with $h^{1,1}(X)=3786$ and $h^{3,1}(X)=2$, which is also in the
database \cite{Kreuzer-Skarke-4d}, but there is no other threefold base besides
$\P^3$ with $h^{1, 1}(B) = 1$, so
there are no other elliptic Calabi-Yau fourfolds with $h^{1, 1}(X) =
2$ besides the generic elliptic fibration over $\P^3$, and the dual
Calabi-Yau in this case it seems is not a simple
elliptic fibration. 

\begin{table}
\centering
\begin{tabular} {|c|c|c|c|c|c|c|c|}
\hline
$h^{1,1}(B)$ (toric) & gauge group & $h^{1,1}(X)$ & $h^{3,1}(X)$ & Mirror base \\
\hline
1943 & $E_8^{29}\times F_4^{81}\times G_2^{216}\times SU(2)^{324}$ & 3277 & 3 & $\tilde{\mathbb{F}}_0$\\
2015 & $E_8^{30}\times F_4^{84}\times G_2^{224}\times SU(2)^{336}$ & 3397 & 3 & $\tilde{\mathbb{F}}_1$\\
2303 & $E_8^{34}\times F_4^{96}\times G_2^{256}\times SU(2)^{384}$ & 3878 & 2 & $\mathbb{P}^3$\\
2591 & $E_8^{38}\times F_4^{108}\times  G_2^{288}\times SU(2)^{432}$ & 4358 & 3 & $\tilde{\mathbb{F}}_3$\\
3086 & $E_8^{45}\times F_4^{129}\times  G_2^{343}\times SU(2)^{513}$ & 5187 & 4 & $\tilde{\mathbb{F}}_4$\\
3686 & $E_8^{54}\times F_4^{153}\times  G_2^{409}\times SU(2)^{615}$ & 6191 & 5 & $\tilde{\mathbb{F}}_5$\\
4373 & $E_8^{64}\times F_4^{180}\times  SO(8)\times G_2^{486}\times SU(2)^{729}$ & 7341 & 7 & $\tilde{\mathbb{F}}_6$\\
5143 & $E_8^{75}\times F_4^{213}\times  G_2^{571}\times SU(2)^{858}$ & 8629 & 7 & $\tilde{\mathbb{F}}_7$\\
5989 & $E_8^{87}\times F_4^{249}\times  G_2^{664}\times SU(2)^{999}$ & 10045 & 7 & $\tilde{\mathbb{F}}_8$\\
10124 & $E_8^{145}\times F_4^{423}\times  G_2^{1125}\times SU(2)^{1683}$ & 16959 & 10 & $\tilde{\mathbb{F}}_{12}$\\
11341 & $E_8^{162}\times F_4^{474}\times  G_2^{1261}\times SU(2)^{1887}$ & 18994 & 12 & $\tilde{\mathbb{F}}_{13}$\\
12631 & $E_8^{180}\times F_4^{528}\times  G_2^{1405}\times SU(2)^{2103}$ & 21151 & 12 & $\tilde{\mathbb{F}}_{14}$\\
\hline
\end{tabular}
\caption[E]{\footnotesize A list of end point bases with the feature that the
  generic elliptic fibration over the base gives a Calabi-Yau fourfold with
  interesting Hodge numbers. In these cases the fourfolds seem to form
 mirror pairs with generic
  elliptic Calabi-Yau fourfolds over simple ``mirror bases'' with
  small $h^{1,1}(B)$. The $h^{1,1}(B)$ listed in the first column
  means the $h^{1,1}(B)$ of the toric base before the codimension-two
  (4,6) loci on divisors with $E_8$ are blown up.} \label{t:endpoints}
\end{table}

It is hard to directly prove that these Calabi-Yau fourfolds with
large $h^{1,1}$ are actually the mirrors of the Calabi-Yau fourfolds
with small $h^{1,1}$ and large $h^{3,1}$, because the Calabi-Yau
fourfolds over specific threefold bases with large $h^{1,1}$ 
are generally hard to realize
explicitly as hypersurfaces in reflexive polytopes. It is natural that
the Calabi-Yau fourfolds with the same Hodge numbers could be
isomorphic to each other. But it is still an open question how to
check this isomorphism.

\subsubsection{Approximating the total number of resolvable and good bases}

Given the results of the 2,000 one-way Monte Carlo runs starting from
$\P^3$, we can try to estimate the total number of resolvable and good
bases
that can be reached as blow-ups of $\P^3$ using
(\ref{dweight}), (\ref{ND}) and (\ref{Ngood}). 

\begin{figure}
\centering
\includegraphics[height=7cm]{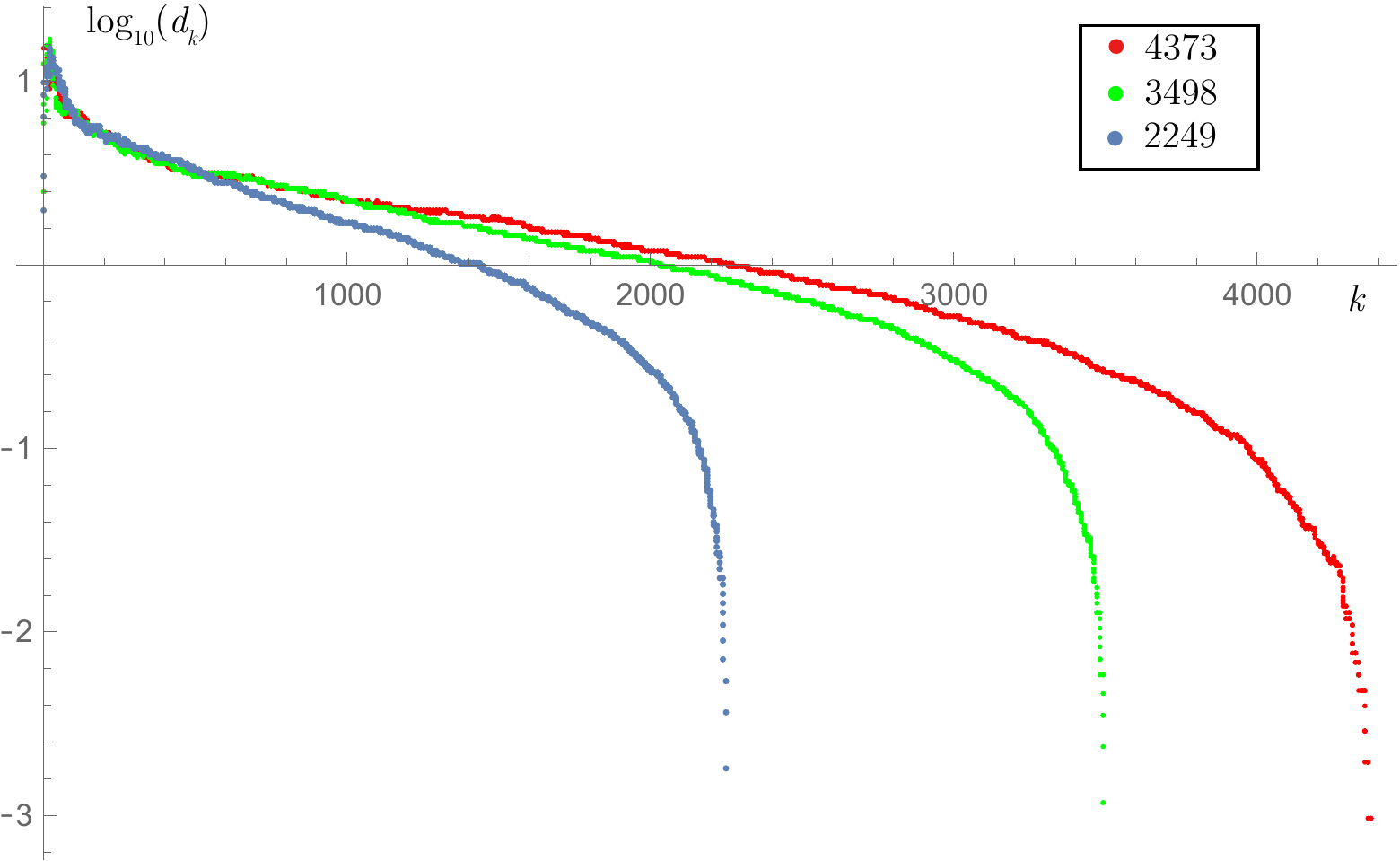}
\caption[E]{\footnotesize The plot of
  $\log_{10}(d_k)=\log_{10}\left(\frac{N_{\rm out}(a_k)}{N_{\rm
      in}(a_{k+1})}\right)$ for each node $a_k$ in each layer $k$ for
  three random blow-up sequences, whose end points are $h^{1,1}(B)=4373$,
  3498 and 2249 respectively.} \label{f:logd}
\end{figure}

For each run of the one-way Monte Carlo, the dynamic weight $D(n)$
given by (\ref{dweight}) gives an estimate of the total number of
accessible nodes at level $n$.  We can write this as
\begin{equation}
D (n) = \prod_{i = 1}^{n-1}  d_k, \;\;\;\;\;
d_k = \frac{N_{\rm out}(a_k)}{N_{\rm in}(a_{k+1})}\,.
\label{eq:weight-product}
\end{equation}
To get a sense of the overall shape of the graph,
we plot in Figure~\ref{f:logd} the value of 
\be
\log_{10}(d_k)\equiv\log_{10}\left(\frac{N_{\rm out}(a_k)}{N_{\rm in}(a_{k+1})}\right)
\ee
at each layer $k$ for three specific random sequences, which end at
$h^{1,1}=4373$, 3498 and 2249 respectively. 
Note that
\begin{equation}
\log_{10} D (n) = \sum_{i = 1}^{n -1} \log_{10} d_k \,.
\end{equation}
One
can see that for each run
there is a specific value $k=k^\star$ where
$\log_{10}(d_k)$ changes sign. That location of $k$ will correspond to
the maximal value of the weight factor $D({k^\star})$ along a particular
trajectory through the graph, which varies with the trajectory.
For small $k$, the weight factor from (\ref{eq:weight-product}) will
grow roughly exponentially, until as $k$ reaches $k^\star$ on a given
trajectory the weight factor peaks and begins to go down.

\begin{figure}
\centering
\includegraphics[height=9cm]{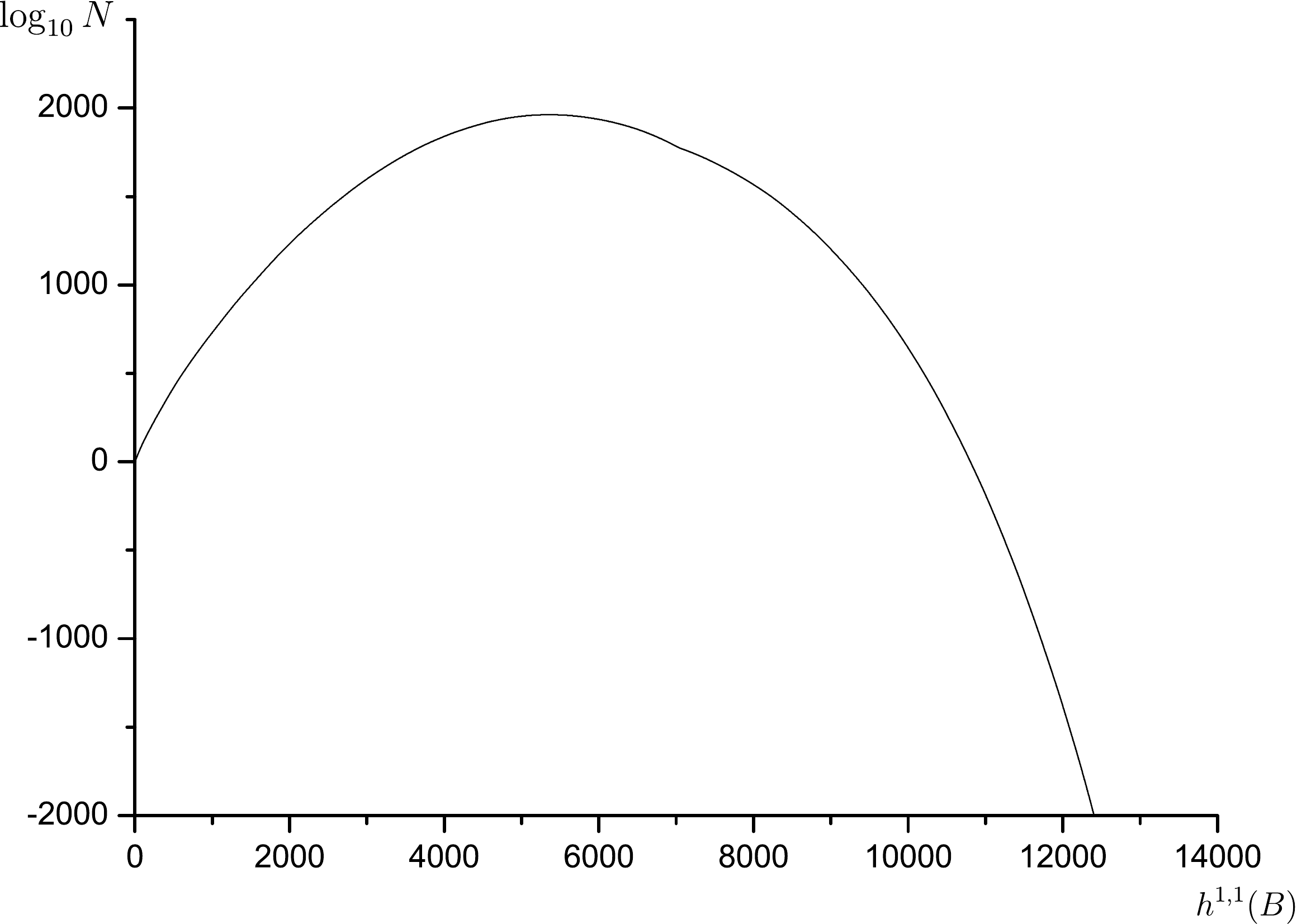}
\caption[E]{\footnotesize Logarithm of the estimated
number of resolvable bases
  $N_{\rm nodes}$ as function of $h^{1,1}(B)$, from blowing up $\mathbb{P}^3$.} \label{f:resP3}
\end{figure}

\begin{figure}
\centering
\includegraphics[height=9cm]{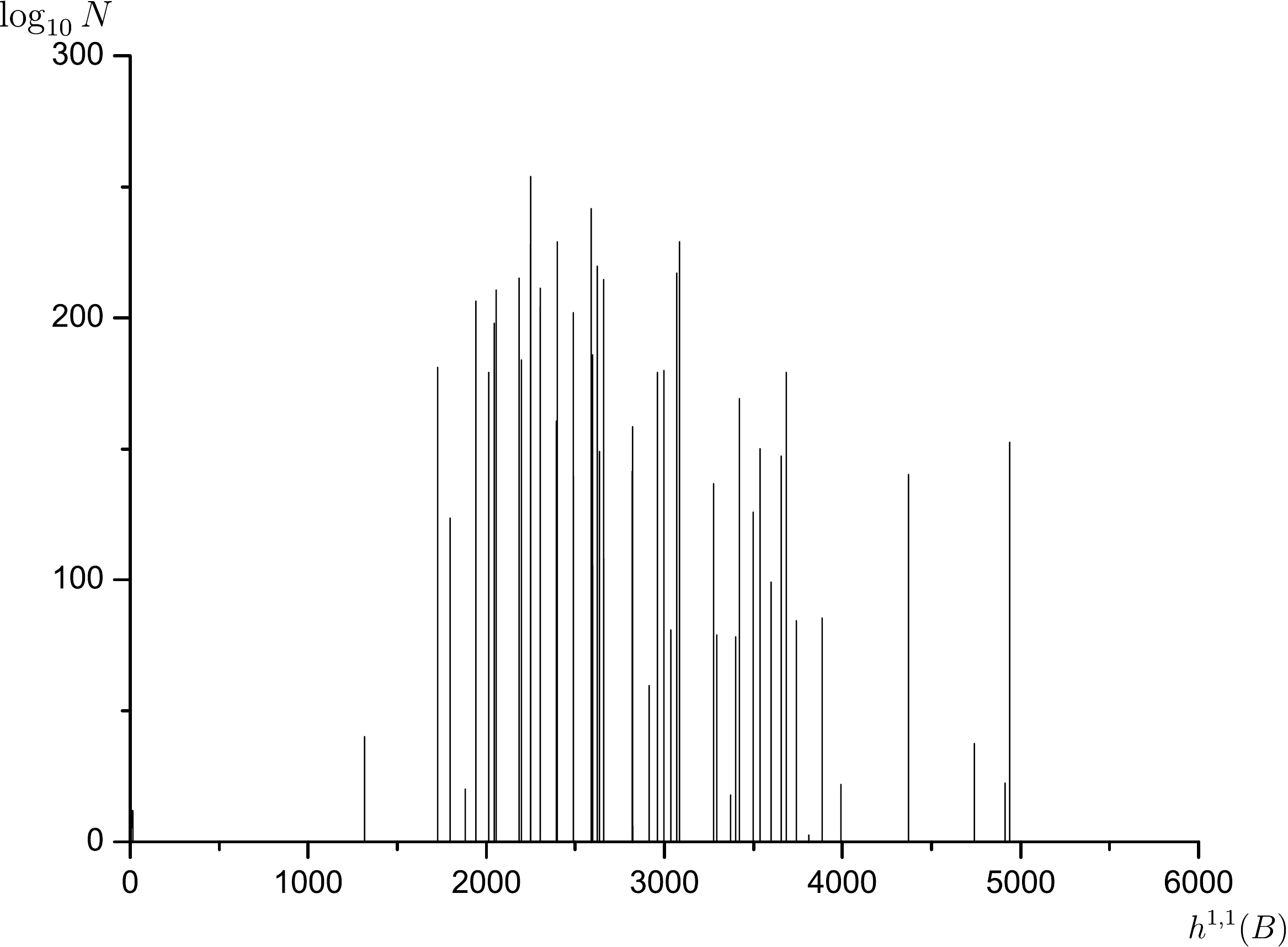}
\caption[E]{\footnotesize Logarithm of the estimated number of good bases $N_{\rm good}$ as function of $h^{1,1}(B)$, from blowing up $\mathbb{P}^3$.} \label{f:goodP3}
\end{figure}

Combining the different runs, we can compute the average value of $D
(n)$ at each $n$ using (\ref{ND}), which gives the expected number of
resolvable bases, and the number of good bases can be estimated as
discussed following equation (\ref{Ngood}) by using (\ref{ND}) with a
0 contribution to $D$ for any trajectory that does not reach a good
base at level $n$.  We plot the logarithm of the estimated
number of resolvable
bases and good bases in Figure~\ref{f:resP3} and Figure~\ref{f:goodP3}
respectively.  As one may expect, the number of resolvable bases
varies smoothly.  As we have discussed above, however, the
distribution of good bases consists of spikes.  Because the weight
factor has a large exponent, and as seen in Figure~\ref{f:logd} the
$k$-dependence of $d_k$ varies fairly widely between runs, the
distribution of resolvable bases is dominated by the trajectory with
the largest value of $k^*$, while the distribution of good bases is a
sum from the spikes of the different end point bases.


We can see from the figure that when $h^{1,1}(B)>5,000$, the estimated
number of good bases with that $h^{1,1}(B)$ is significantly smaller
than 1, even though we found some good end point bases with much
larger values of $h^{1, 1}(B)$. Moreover, the estimated number of
resolvable bases is smaller than 1 at $h^{1,1}(B)>11,000$. For
example, for the base with biggest $h^{1,1}(B)=12,631$, the total
number of bases is estimated as $2.2\times 10^{-2474}$. This fact
verifies that we have indeed underestimated the total number of bases
by a large exponential factor. Typically there are $\sim 10^3$
incoming edges for the bases near the end points but only a few
outgoing edges, hence the situation is more extreme than
Figure~\ref{f:strange} shows.

Using the uncorrected $N_{\rm in}(a_k)$ in (\ref{dweight}), the
estimated total number of resolvable bases is equal to $3.5\times
10^{1964}$ and the estimated number of good bases equals to $3.0\times
10^{253}$.  If we use the corrected estimation $\tilde{N}'_{\rm
  in}(a_k)$ as discussed in Section \ref{sec:systematic-issues}, then
the total number of resolvable bases is again estimated at to
$3.5\times 10^{1964}$ and the number of good bases is estimated at
$9.1\times 10^{253}$. Hence from the experimental results, it seems
that the different definition between $\tilde{N}'_{\rm in}(a_k)$ and
$N_{\rm in}(a_k)$ does not much affect the estimation, so this
mechanism does not capture the source of the underestimation.
We discuss the reasons for the underestimation and possible
corrections further in Section \ref{sec:error}.

Note that even with the underestimation error, it is clear that the
set of resolvable bases that can be constructed through this approach
is significantly larger than the $10^{755}$ bases found in
\cite{Halverson:2017}.  We can relate these computations in terms of
the possible ``heights'' of divisors $h(D_i)$ used in that paper. We
assign a number 1 to each divisor on the starting point
$\mathbb{P}^3$. Whenever we blow up a 3D cone $D_i D_j D_k$, we assign
a number $h(E)=h(D_i)+h(D_j)+h(D_k)$ to the exceptional
divisor. Similarly, if we blow up a 2D cone $D_i D_j$, we assign a
number $h(E)=h(D_i)+h(D_j)$ to the exceptional divisor. In
\cite{Halverson:2017} they use a sufficient criterion that the height
of any divisor on a base cannot exceed 6 in order to avoid
codimension-one (4,6) loci. However, in our Monte Carlo algorithm, the
typical maximal height of a divisor on an end point ranges from
50--350, which drastically exceeds the limit 6.

\subsection{Blowing up other starting points with small $h^{1,1}(B)$}

To cross-check the results in the previous section, we have also
looked at blow-up trajectories
from other starting point bases with small $h^{1,1}(B)$.

Two other starting bases we have tried are the generalized Hirzebruch
threefold $\tilde{\mathbb{F}}_2$ with $h^{1,1}(B)=2$ and a simple
product space $\mathbb{P}^1\times\mathbb{P}^1\times\mathbb{P}^1$ with
$h^{1,1}(B)=3$. 
Similar
to $\mathbb{P}^3$, these bases do not have non-Higgsable gauge
groups. 
After 1,000 random blow up sequences starting from
$\tilde{\mathbb{F}}_2$, we found a larger fraction of end point
bases with large $h^{1,1}(B)$ than when starting from $\P^3$. 
For $\tilde{\mathbb{F}}_2$,
1\% of end points have $h^{1,1}(B)>10,000$,
while this percentage is 0.3\% from the starting point
$\mathbb{P}^3$. The largest $h^{1,1}(B)$ we got is 20,341, and the
non-Higgsable gauge group on the resulting good endpoint
base is $E_8^{290}\times
F_4^{850}\times G_2^{2261}\times SU(2)^{3383}$. We estimate the total
number of resolvable bases from $\tilde{\mathbb{F}}_2$ at $1.24\times
10^{3046}$ while the total number of good bases is estimated at
$1.10\times
10^{254}$, using weight factors with $N'_{\rm in}(a_k)$. On the other
hand, after 1,000 random blow up sequences from
$\mathbb{P}^1\times\mathbb{P}^1\times\mathbb{P}^1$, the total number
of resolvable bases is estimated to be $1.43\times 10^{1811}$ and the
total number of good bases is estimated to be $1.80\times 10^{271}$.

\begin{figure}
\centering
\includegraphics[height=9cm]{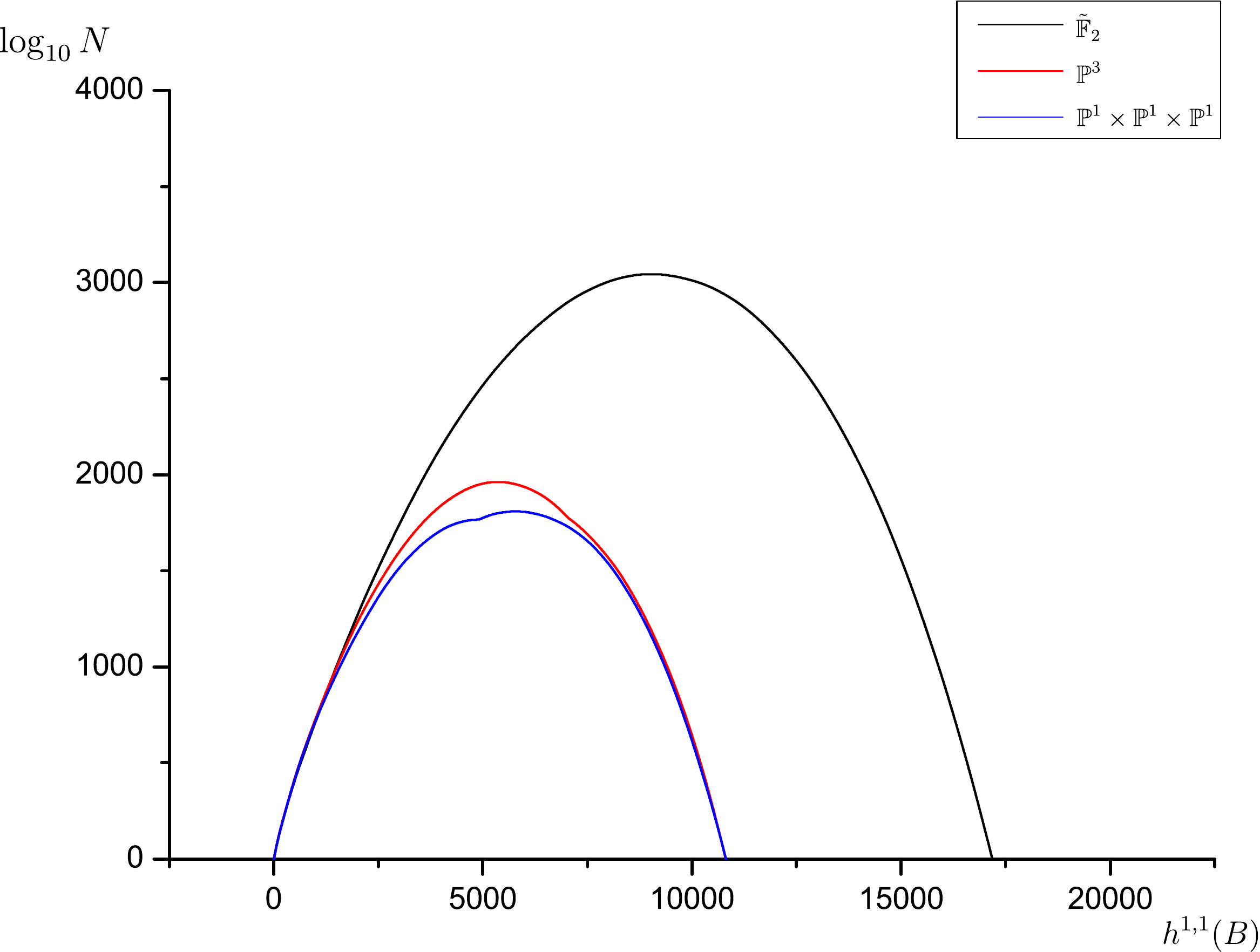}
\caption[E]{\footnotesize Logarithm of the estimated number of resolvable bases as function of $h^{1,1}(B)$ from blowing up $\mathbb{P}^3$, $\tilde{\mathbb{F}}_2$ and $\mathbb{P}^1\times\mathbb{P}^1\times\mathbb{P}^1$.} \label{f:res}
\end{figure}

\begin{figure}
\centering
\includegraphics[height=9cm]{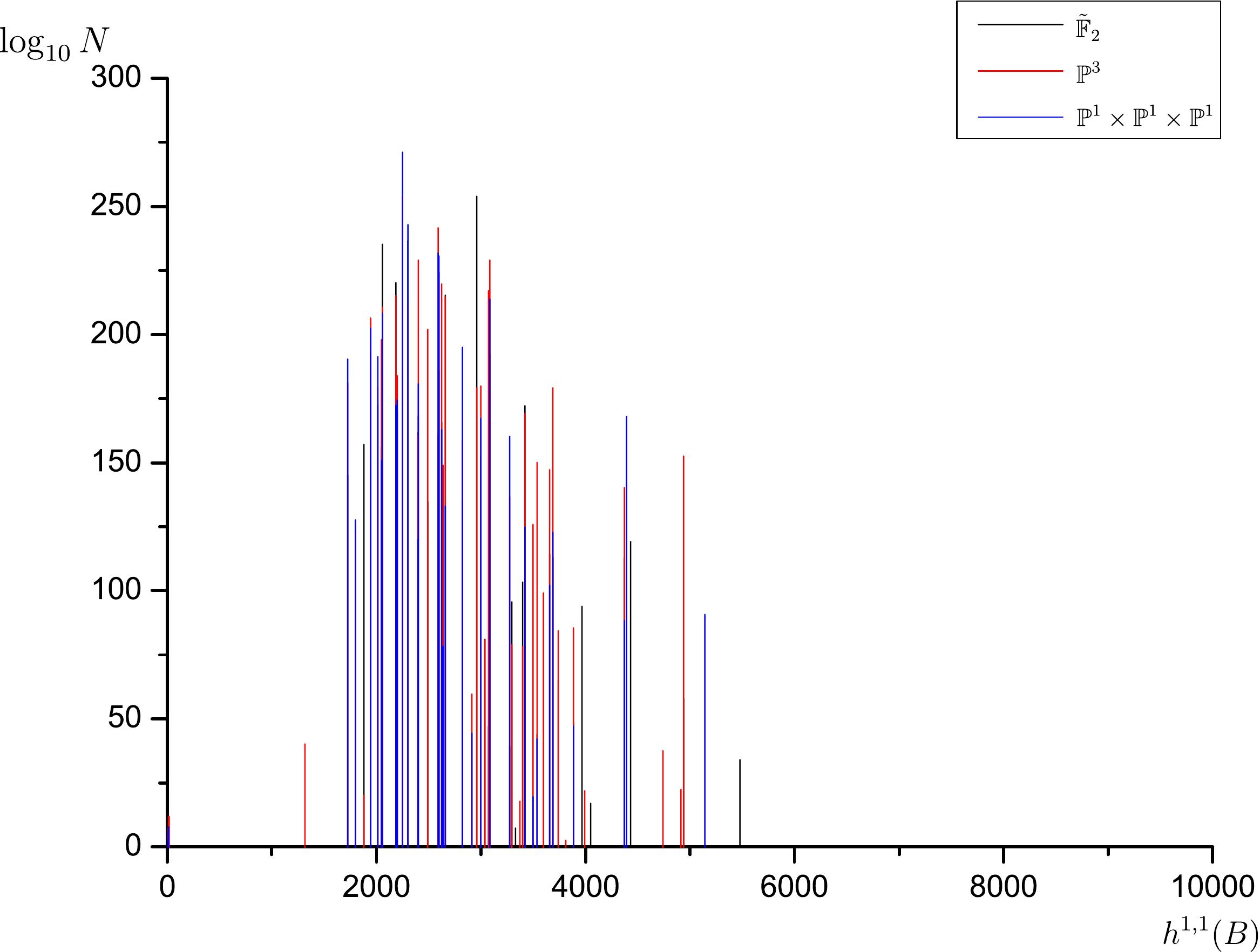}
\caption[E]{\footnotesize Logarithm of the estimated
number of good bases as function of $h^{1,1}(B)$  from blowing up $\mathbb{P}^3$, $\tilde{\mathbb{F}}_2$ and $\mathbb{P}^1\times\mathbb{P}^1\times\mathbb{P}^1$.} \label{f:good}
\end{figure}

We plot the distribution of resolvable bases and good bases from the three starting points in Figures~\ref{f:res} and ~\ref{f:good} respectively.

Many peaks in Figure~\ref{f:good} indeed overlap between starting
points; for example, the total number of good bases with
$h^{1,1}(B)=2249$, 2303, 2591, 2961 and 2999 for the starting points
($\mathbb{P}^3$,$\mathbb{F}_2$,$\mathbb{P}^1\times\mathbb{P}^1\times\mathbb{P}^1$)
are on the order of $(10^{254},10^{208},10^{271})$,
$(10^{211},10^{236},10^{243})$, $(10^{241},10^{161},10^{232})$,
$(10^{179},10^{254},10^{-227})$ and
$(10^{180},10^{152},10^{167})$ respectively. The highest peaks from
the starting points
($\mathbb{P}^3$,$\mathbb{F}_2$,$\mathbb{P}^1\times\mathbb{P}^1\times\mathbb{P}^1$)
locate at $h^{1,1}(B)=(2249, 2961, 2249)$ respectively. Because of the
big exponential error on these estimations, we cannot fix the exact
peak in the distribution of good bases, and these numbers are clearly
not very accurate estimates.
We discuss some related issues further in Section~\ref{sec:global}

Similar to the runs from $\mathbb{P}^3$, the total number of bases
with large $h^{1,1}(B)$ seems to be hugely underestimated. Starting
from $\tilde{\mathbb{F}}_2$, we get an extremely small estimation
$6\times 10^{-4221}$ for the number of bases with
$h^{1,1}(B)=20,341$. Starting from
$\mathbb{P}^1\times\mathbb{P}^1\times\mathbb{P}^1$, we find that the
total number of bases with $h^{1,1}(B)=12,631$ is estimated at
$5\times 10^{-2454}$.
Since in both cases the actual number must be at least 1, we may
roughly expect that the error in the exponent grows linearly with $k$,
so that the true number of bases may be closer to $10^{6000}$ than the
underestimate $10^{3000}$, though more careful analysis would be
needed to nail down this number more precisely.

As another starting point with somewhat different structure, we have
also
tried blowing up starting with the generalized Hirzebruch threefold
$\tilde{\mathbb{F}}_{12}$, which has $h^{1,1}(B)=2$ and a non-Higgsable
gauge group $E_7$. Despite the existence of an $E_7$, the Weierstrass
model does not suffer from the problem of codimension-three (4,6)
loci and non-flat fibration described in Section~\ref{sec:toricF},
because the coefficient $b_0$ in (\ref{E7model}) cannot vanish. After
100 random blow up sequences, the total number of resolvable bases is
estimated as $1.77\times 10^{2130}$ while the total number of good
bases is estimated as $2.02\times 10^7$.

The distribution of $h^{1,1}(B)$ among the 100 runs starting from
$\tilde{\mathbb{F}}_{12}$ has a greater preference towards large
$h^{1,1}(B)$ than the other starting points we have investigated. 17\%
of end point bases have $h^{1,1}(B)>10,000$, and the largest
$h^{1,1}(B)$ we found is $33,021$. Nonetheless, we can still find 1
base with $h^{1,1}(B)=2249$, 4 bases with $h^{1,1}(B)=2303$ and 8
bases with $h^{1,1}(B)=2591$, which are common bases in
Table~\ref{t:distribution}. Another interesting phenomenon is that
none of the end point bases has an $E_7$ gauge group factor. The divisor
possessing an $E_7$ gauge group at the starting point always has an
$E_8$ at the end point instead.

\subsection{Blowing up the base $B_{\rm max}$
that gives rise to the elliptic CY fourfold $\mc{M}_{\rm max}$ with the largest $h^{3,1}$}

In \cite{Maxh31}, we explicitly constructed the base $B_{\rm max}$
such that the generic elliptic CY fourfold $\mc{M}_{\rm max}$ over it
has the largest $h^{3,1}$: $(h^{1,1},h^{3,1})=(252,303148)$.  This is
the largest value of $h^{3,1}$ known for any Calabi-Yau fourfold,
elliptic or not.  As argued in \cite{Maxh31}, we believe this should
be the largest possible value for any elliptic Calabi-Yau fourfold,
but unlike the analogous maximum value of $h^{2, 1}(X_3) = 491$ for
elliptic Calabi-Yau threefolds \cite{Hodge}, we do not have a rigorous
proof of the upper bound for fourfolds.  The base $B_{\rm max}$ is
constructed as a $B_2$ bundle over $\mathbb{P}^1$, where $B_2$ is a
toric surface characterized by a closed cycle of toric curves with
self-intersections $0, + 6,$
-12//-11//-12//-12//-12//-12//-12//-12//-12, where $//$ denotes the
``$E_8$ chain'' of self-intersections $-1, -2, -2, -3, -1, -5, -1, -3,
-2, -2, -1$.  The generic elliptic CY threefold over $B_2$ is a
self-mirror CY3 with Hodge numbers $(251, 251)$ \cite{mt-toric, Hodge}.
The toric rays $w_i$ on $B_2$ can be choosen to be
\begin{eqnarray}
w_1 & = &  (-1, -12)\\
w_2 & = &  (0, 1)\\
w_3 & = & (1, 6)\\
\vdots & & \vdots\\
w_{99} & = & (0, -1) \,.
\end{eqnarray}
The intermediate rays can be determined by the condition
$w_{i -1} + w_{i + 1}+ (C_i \cdot C_i) w_i = 0$, where $C_i \cdot C_i$
is the self-intersection of the $i$th curve.
From the rays $w_i$ we can construct the toric fan for $B_{\rm max}$, which
is given by the rays
\begin{eqnarray}
v_0 & = & (0, 0, 1)\\
v_i & = & (w_i, 0), \;\;\;\;\; 1 \leq i \leq 99\\
v_{100} & = & (84, 492, -1) = (12w_{15}, -1) \,,
\end{eqnarray}
where $C_{15}$ is the curve in $B_2$ of self-intersection $-11$.  The
3D cones of the fan for $B_{\rm max}$ are given by $(v_0, v_i, v_{i + 1})$
and $(v_{100}, v_i, v_{i + 1})$, including the ones 
$(v_0/v_{100}, v_{99}, v_1)$.

We performed 100 random blow up sequences from $B_{\rm max}$, getting
100 endpoints with $h^{1,1}(B)=2636-16103$. The total number of
resolvable bases is estimated at $1.14\times 10^{520}$ while the total
number of good bases is $4.3\times 10^4$, which is an extremely small
number. This is a common feature of this Monte Carlo approach when one
starts from bases with large non-Higgsable gauge groups, and we can
conclude that the number of bases we can get by blowing up bases with
larger $h^{1,1}(B)$ is significantly smaller than the number of bases
we can get by blowing up $\mathbb{P}^3$.  This matches with what we
expect from the case of base surfaces where the complete set of toric
bases is known.

It is notable that although the structure of $B_{\rm max}$ looks
completely different from $\mathbb{P}^3$, 42\% of the end point bases
we get from blowing up $B_{\rm max}$ have $h^{1,1}(B)$ that can be
found in Table~\ref{t:distribution}. For example, 7\% of bases have
$h^{1,1}(B)=7909$ and 5\% of bases have $h^{1,1}(B)=8980$. This
implies that the end point bases from different starting points have a
large overlap, and the ensemble of bases we get from blowing up
$\mathbb{P}^3$ may be a decent 0th order approximation to the total
set of 3D toric bases.

Another interesting challenge is to find a blow-up sequence to
explicitly generate the base $B$ with the largest value of
$h^{1,1}(B)$, which gives rise to the elliptic fourfold $X$ with the
largest $h^{1,1}(X)$. In the lower dimensional (base surface) case, we
know that the generic elliptic Calabi-Yau threefold over
$\mathbb{F}_{12}$ has Hodge numbers $(h^{1,1},h^{2,1})=(11,491)$. If
one blows it up, the ``end point'' bases include one with
$h^{1,1}(B)=174$, which gives rise to the Calabi-Yau threefold with
$(h^{1,1},h^{2,1})=(491,11)$. These two Calabi-Yau threefolds are the
currently known (irreducible) Calabi-Yau threefolds with the largest
$h^{2,1}$ and $h^{1,1}$ respectively. So if we start from the base
$B_{\rm max}$ and blow it up, requiring that the number of complex
structure moduli is maximized at each step, we may expect to finally
reach a base with the largest $h^{1,1}(B)$ that gives rise to the
mirror fourfold of $\mc{M}_{\rm max}$, with
$(h^{1,1},h^{3,1})=(303148,252)$ \cite{Klemm-lry}. But it is quite
time consuming to actually perform such a search, since one has to
enter a region where typical bases have $\mc{O}(100,000)$ rays and
$\mc{O}(100,000)$ complex structure moduli; we leave the resolution of
this challenge to future work.

\subsection{Systematic error of the estimated total number}
\label{sec:error}

As we have mentioned before, the number of bases at large
$h^{1,1}(B)$ tends to be substantially
underestimated by using the weight factor  through
(\ref{ND}). 
For example, the weight factor suggests that for the starting point 
 $\tilde{\mathbb{F}}_2$, the total number of bases with $h^{1, 1}(B) >
  20,000$ is something like $10^{-4000}$.  Not only do we know that
  there is at least one base in this range, but there is actually a
  large multiplicity of such bases.

\begin{figure}
\centering
\includegraphics[height=7cm]{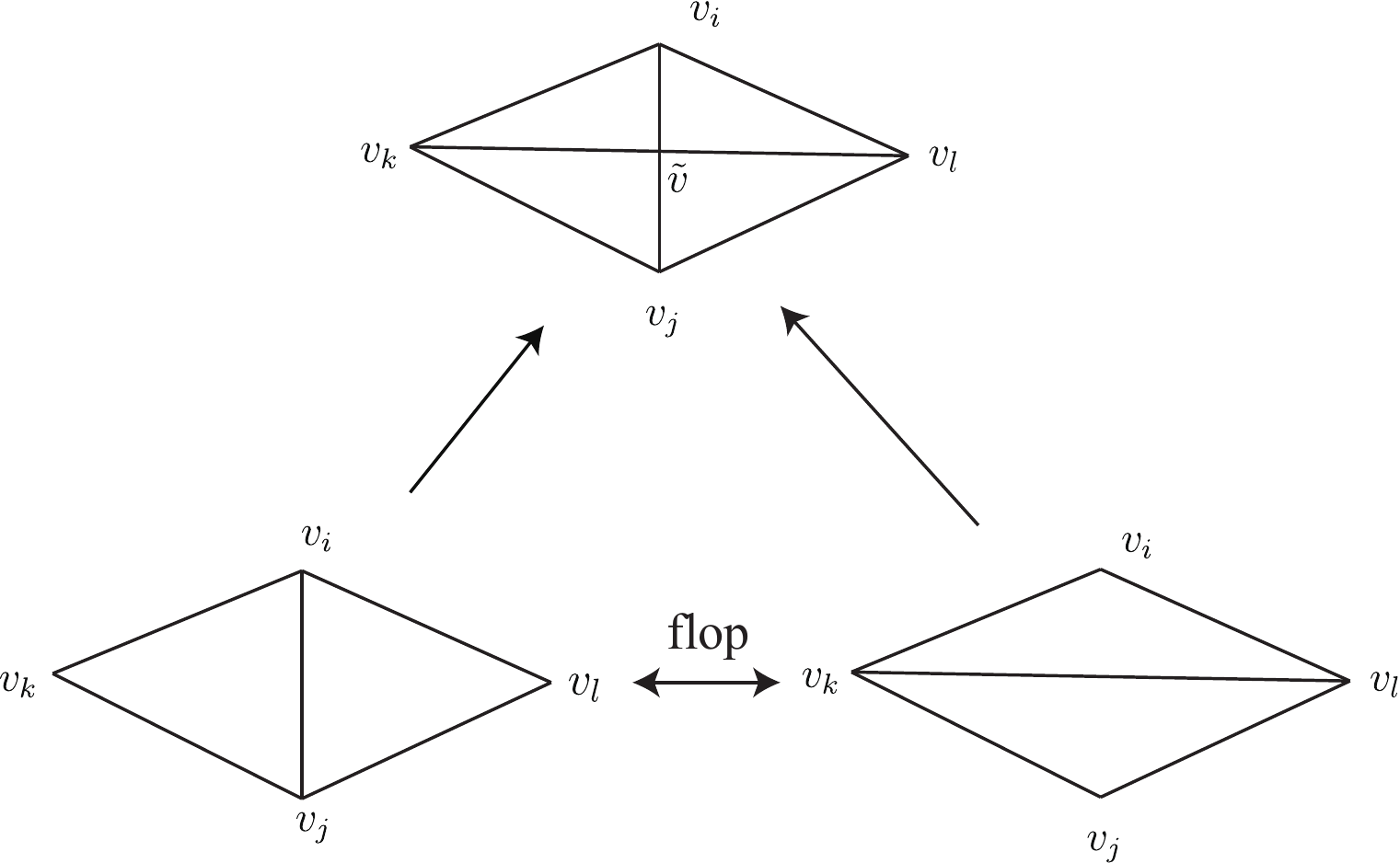}
\caption[x]{\footnotesize An illustration of 
the flop process, which can happen when $v_i+v_j=v_k+v_l$.}\label{f:flop}
\end{figure}

We can get a rough estimate of the multiplicity of bases at the large
Hodge numbers where there are good end points by considering flops.
In general, for many of the bases we encounter we know that there
exist a large number of flop operations on the base which transform it
to another base with the same rays but a different cone structure.
Such a flop appears when there are rays $v_i,v_j,v_k,v_l$ satisfying
the relation $v_i+v_j=v_k+v_l$, and there is a 2d cone $v_iv_j$, see
Figure~\ref{f:flop}.  When there are many rays on a face of the convex
hull of the polytope formed from the base fan, there are generally
many possible flops.  For example, the number of possible flop
operations on an end point base with $h^{1,1}(B)=10,124$ ranges from
3440 to 3980, and there are 7040 flops on an end point base with
$h^{1,1}(B)=20,341$. In general, we can empirically estimate that the
number of flops on an end point base scales linearly as
$N_{flop}\approx 0.3 h^{1,1}(B)$--$0.4h^{1,1}(B)$.  As discussed in
\cite{MC}, a rough estimate for the number of distinct bases
associated with a given base with $n$ flops is $2^n$.  This implies
that there are at least something like $\sim 2^{3500}\approx
10^{1000}$ bases with $h^{1,1}(B)=10,124$ and at least $\sim
2^{7000}\approx 10^{2000}$ bases with $h^{1,1}(B)=20,341$, just from
the different combinations of flop operations on these bases.  In the
set of bases studied in \cite{MC}, flops were not the primary source
of multiplicity for bases with common Hodge numbers.  Indeed, for the
good end point base values of $h^{1, 1}(B)$ encountered here, there
are generally many bases with different set of rays that cannot be
identified with flops.  Thus, the number of different end point bases
with $h^{1,1}(B)\gtrsim 10,000$ should be significantly larger than 1.

The underestimation of the number of bases at large $h^{1, 1}(B)$ has
two potential origins that we are aware of:

(1) The branches with highest probability will have a weight factor
that is lower than the actual average weight factor, leading to a
systematic exponential underestimate of the number of bases for
typical blow-up trajectories. 
As in the toy model described in Appendix A,
this effect can also produce an exponential suppression $e^{-ak}$ on
the weight factor.

(2) We are overcounting $N_{\rm in}$ because for some of the bases $B$
that are generated by blowing up the starting point a number of times
some of the incoming edges are associated with contractions to bases
that cannot be contracted to the starting point through a sequence of
smooth bases. This overcounting issue may occur at most levels of the
blow-up process, which is likely to lead to a roughly exponential
underestimation $e^{-bk}$ on the weight factor $D$.

\vspace*{0.05in}
Because both of these effects can have an exponential suppression of
the weight factor it is important to distinguish their relative significance.
We believe that the
factor (2) is the primary cause of the underestimation, but we cannot
quantify it precisely yet since it is difficult to tell whether a base can be
contracted to $\mathbb{P}^3$ or not.
This issue may be addressed by systematically understanding what
fraction of incoming edges $N_{\rm in}$ can lead to a contraction back
to the starting point, or potentially by systematically understanding
the set of starting point bases that are either smooth or admit
certain kinds of singularities.  We discuss the two possible issues
related to undercounting in turn, first by   arguing that (1) is not the principal source of
underestimation error, and then by
making some simple
observations on the existence of additional starting points relevant
to (2).

\subsubsection{Weight factors and the lognormal distribution}

Problem (1) is illustrated with a toy model of a homogeneous random
graph in Appendix A, where the actual number of nodes for each layer
$k$ is constant, but the formula (\ref{ND}) will lead to an
exponentially small number $D\propto e^{-ak}$ for typical blow-up
trajectories, and the proper average is only restored by sampling
exponentially unlikely trajectories.  In this simple model, the
logarithm of $D (k)/D (k -1)$ is essentially a random variable sampled
independently at each level, so the
distribution of weight factors $D$ at layer $k$ obeys a lognormal
distribution and the standard deviation $\sigma$ of $\log_{10}(D)$
scales as $\sigma\propto\sqrt{k}$, while the mean
$\mu$ of $\log_{10}(D)$ scales as $k$.  In such a situation, we can
compute the expectation value of the weight factor:
\be \bsp \langle D\rangle &=\int_{-\infty}^\infty
e^{x\log
  10}\cdot\frac{1}{\sqrt{2\pi\sigma^2}}e^{-\frac{(x-\mu)^2}{\sigma^2}}dx\\ &=\exp\left[\mu\log
  10+\frac{\sigma^2(\log 10)^2}{2}\right].\label{Expwf}
\end{split}
\ee
Typical trajectories will only see the first term, so there is a
systematic underestimate of $D$ of the form $e^{-ak}$ that can be
corrected by including the
$\sigma^2$ term in the exponent by hand as a correction factor.

\begin{figure}
\centering
\includegraphics[height=5cm]{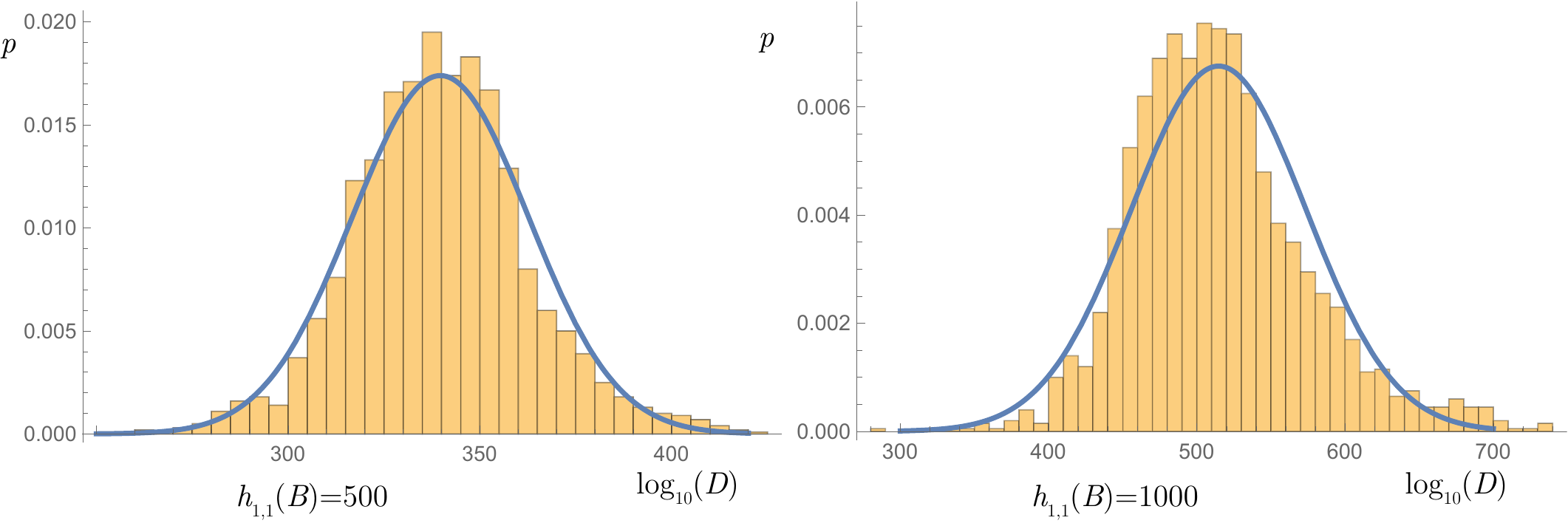}
\caption[E]{\footnotesize Distribution of the logarithm of the dynamic weight factor $D$ at $h^{1,1}(B)=500$ and 1,000 and the fitting curve using a normal distribution. For $h^{1,1}(B)=500$, the mean value of $\log_{10}(D)$ is 339.707 while the standard deviation is 22.9357. For $h^{1,1}(B)=1,000$, the mean value of $\log_{10}(D)$ is 514.694 while the standard deviation is 59.0485.} \label{f:weightfactor}
\end{figure}

While this analysis is directly applicable for homogeneous graphs  like the toy
model in Appendix A, which have a structure that is identical at each
level and homogeneous across the graph at each level, it is more subtle to use this
kind of analysis for the graph of toric threefold bases sampled
through successive blow-ups, which is different at each level and not
homogeneous across trajectories that explore different parts of the
graph.  As a simple illustration of this distinction,
we plot the distribution of the logarithm of the weight factors at
$h^{1,1}(B)=500$ and 1,000 starting from $\mathbb{P}^3$ in
Figure~\ref{f:weightfactor}. 
We also fit the logarithm of weight
factors $x=\log_{10}(D)$ using a normal distribution with mean value
$\mu$ and standard deviation $\sigma$. 
It turns out that for this data, the standard deviation grows faster
than $\sqrt{k}$, at least in some regime of  values of $k$.
 Hence
  the actual distribution of $D$ cannot be approximated by the 
simple lognormal distribution arising from a product of equally
distributed factors, as it can for the
  random graph toy model in the Appendix A. 

\begin{figure}
\centering
\includegraphics[height=8cm]{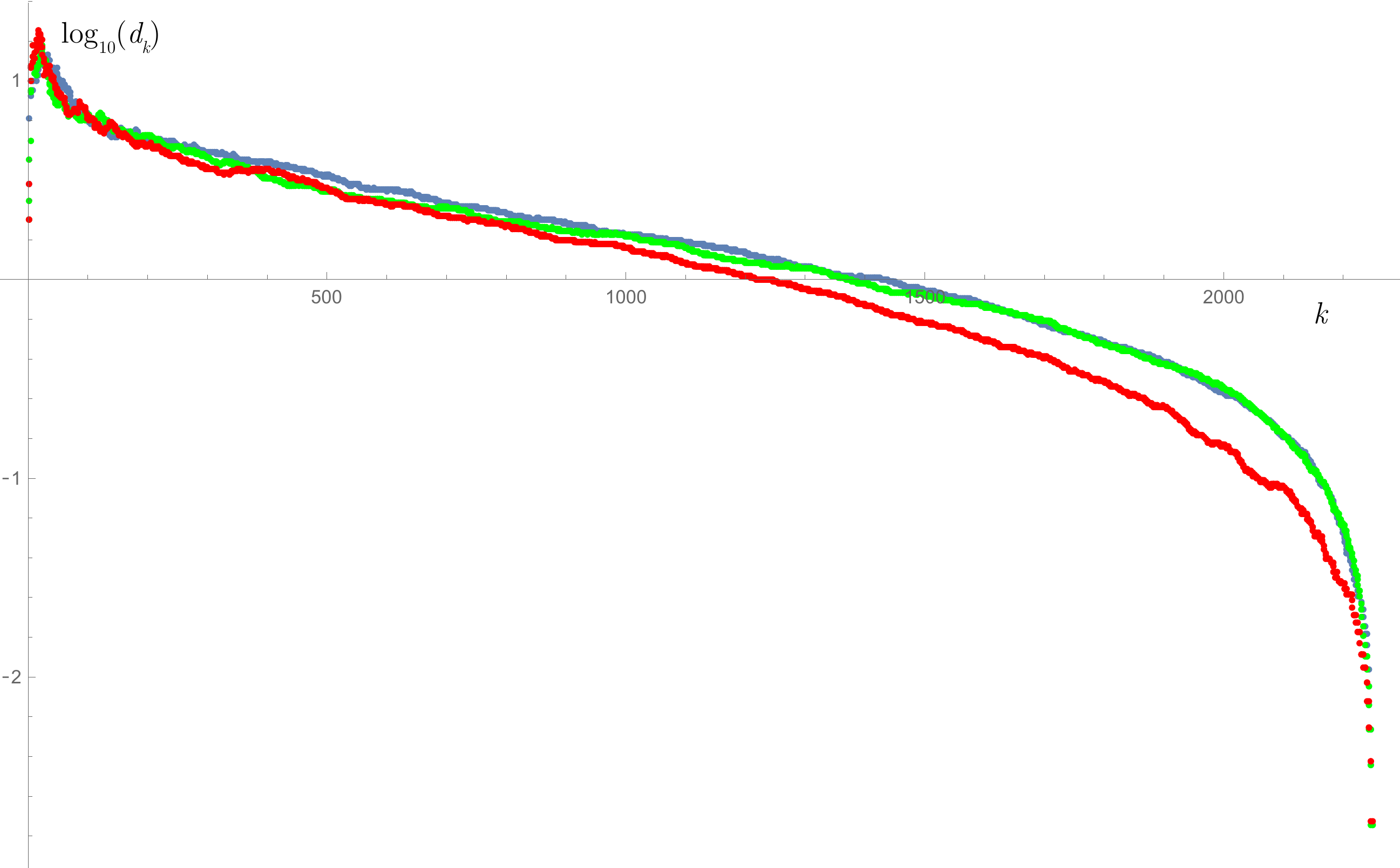}
\caption[E]{\footnotesize The plot of $\log_{10}(d_k)=\log_{10}\left(\frac{N_{\rm out}(a_k)}{N_{\rm in}(a_{k+1})}\right)$ for each node $a_k$ in each layer $k$ for three random blow up sequences, whose end points are bases with $h^{1,1}(B)=2249$.} \label{f:logdg}
\end{figure}

An important distinction in the toric threefold blow-up graph is that
each random blow-up sequence actually enters a separate region of the
graph with different local statistics at a given level. This statement
even holds for different branches that lead to the same end
point. For example, we plot
$\log_{10}(d_k)=\log_{10}\left(\frac{N_{\rm out}(a_k)}{N_{\rm
    in}(a_{k+1})}\right)$ of three different branches that lead to end
points with $h^{1,1}(B)=2249$ in Figure~\ref{f:logdg}. Since
$D(a_k)=\prod_{i=1}^{k-1} d_i$, the weight factor $D$ for the red
branch at $k>1000$ will be significantly smaller than the $D$ for blue
  and green branches.
Furthermore, the divergence between different branches, particularly
those with different endpoints, leads to an extremely large standard
deviation on $D$ when compared across paths.  This would lead to an
unreasonably large correction factor $e^{c \sigma^2}$ from the second
term in the exponent of  (\ref{Expwf}) if we naively assume a
lognormal distribution.

Thus, it does not make sense to consider the full set of trajectories
in the context of a lognormal distribution on $D$.  Nonetheless, we
may imagine that each trajectory follows a path where the local factor
$d_k$ is chosen at each $k$ from some distribution such that the final
distribution on $\log (D)$ can be described as the summation of the
distributions on $\log (d_k)$, which will again give a roughly
lognormal form given the local distribution of $d_k$ factors in each
region of the full graph traversed by a given trajectory.  To estimate
the effect in such a local model, we can compute the local variation
or smoothness of the curve $f(k)\equiv \log_{10}(d_k)$ for a single
branch to roughly estimate the standard deviation responsible for  the
 underestimation of type (1)  along a particular trajectory. 
As described in Appendix A, we can compute the ``local'' standard
deviation
\be
\sigma(a_k)=\sqrt{\frac{\sum_{i=k-l}^{k+l}\left(f(i)-\frac{1}{2l+1}\sum_{j=k-l}^{k+l}f(j)\right)^2}{2l+1}}
\ee
for each node $a_k$ and take the the standard deviation $\sigma$ in the compensation formula (\ref{Expwf}) to be
\be
\sigma=\sqrt{\sum_{i=l+1}^{k-l}\sigma^2(a_i)}.
\ee
Here $l$ can be an arbitrary positive integer with the condition $l\ll
k$.
A slightly more sophisticated analysis could take account of
correlations between successive values of $f (k)$, but this would only
serve to reduce the $\sigma^2$-based correction factor.

As an example of this local analysis, for the blue curve in
Figure~\ref{f:logdg}, if we take $l=4$, then the final standard
deviation for $k=2249$ is $\sigma=0.683$, which is extremely small,
and is essentially a negligible correction to the weight factor given
the large apparent undercounting error in the exponent. 
This can be compared, for example, with 
the toy model in Appendix A, where there is an
underestimation factor of $10^{-50}$ at $k=1000$, and the standard
deviation is $\sigma=21.7307$. This analysis suggests that the unevenness
of the weight factor due to local fluctuations in the structure of the
graph in the vicinity of specific blow-up trajectories,
i.e. explanation (1) above,
is not the major cause of the exponentially small
number $N_{\rm node}\propto e^{-ak}$.

\subsubsection{Exotic smooth starting points}

To investigate the question of whether some of the $N_{\rm in}$
incoming edges at a typical node $B$ may not lead down to a base that can
be contracted to the starting point, a natural question that we can
address is: what is the maximal number of blow downs we can perform
with a random sequence starting at an end point base $B$ before
reaching a starting point that cannot be contracted through a blow
down to another smooth base.  It turns out that after a random
sequence of blow downs on a typical good end point base, we almost
always hit a base with $h^{1,1}(B)=50$--$200$ that cannot be
contracted to another smooth base, even if we require that the
original rays on the starting point are never removed. These ``exotic
starting points'' have a complicated fan structure. Most of the rays
in the fan have more than four neighbors, hence this base is clearly
neither a $\mathbb{P}^1$ bundle over $B_2$ or a $B_2$ bundle over
$\mathbb{P}^1$. An exotic starting point base typically has a lot of
toric curves where $f$ and $g$ vanish to order (4,6) or higher. We
explicitly write down an example of an exotic starting point in
Appendix B.

This observation suggests that there is a large number of such exotic
starting points, however at this point we neither have a good
estimation of their total numbers or their general structures. If we
want to extensively survey the set of resolvable bases, then these
starting points should be taken into account.  Knowing the
distribution of these exotic starting points would also potentially
help in resolving the underestimation issue.  
One possible approach would be to allow for singular bases, as
suggested by Mori theory.  These exotic starting points could be blown
down further by allowing the contraction of rays associated with
divisors other than $\P^2$, giving singular starting points.
However, 
if we only
focus on identifying the set of good bases, then these other starting
points may not be too important since a random blow up sequence from
an exotic starting point may end on a similar class of end point bases
to those encountered here, as we found in the previous subsection
where the same end point values of $h^{1, 1}(B)$ were encountered from
quite different starting points.

\section{Global structure of the set of good bases}
\label{sec:global}

One aspect of the set of toric threefold bases that we would like to
understand better is the distribution and nature of the good bases
without codimension-two (4, 6) curves among the much larger set of
resolvable bases.  In the Monte Carlo experiments we have done, from all the random blow up sequences we
never found any other good bases between $h^{1,1}(B)=20$ and the end
points.  As discussed in the previous section, the good end point
bases have some interesting features; in particular, they seem to lead
to elliptic Calabi-Yau fourfolds with mirror duals that are often
elliptic fibrations over simple bases.  These are not, however, the
only good bases among the toric threefolds.  They are just the only
ones we encounter in the one-way Monte Carlo since (4, 6) curves are
so plentiful that they generally dominate the geometry of the base up
to the final end point of the blow-up sequence.  In this section we
discuss another set of good bases that can be accessed using the
results of our Monte Carlo computations.

Another way to construct a good base besides the end point ones is to
start from some base in the middle of the blow-up sequence and from
there only blow up the codimension-two (4,6) loci wherever they
appear. In this process, many good bases with $h^{1,1}(B)$ different
from the list in Table~\ref{t:distribution} can be generated.  The
non-Higgsable gauge group structure of this type of intermediate good
base (which we denote by $B_{\rm int}$) is similar to the end points,
and the number of each gauge group factor SU(2), $G_2$, $F_4$ and $E_8$ can
be approximated by the formula (\ref{Ngaugegroups}).  One qualitative
difference between these intermediate good bases and the end point
bases is that the generic elliptic fourfold $X$ over a base $B_{\rm
  int}$ typically has a larger $h^{3,1}(X)$, since we prefer blowing
up codimension-two (4,6) loci, which does not reduce the number of
Weierstrass monomials in $f$ and $g$. For example, we can get an $X$
with $h^{1,1}(X)=7097$ and $h^{3,1}(X)=1452$.  For intermediate bases
with larger $h^{3,1}(X)$ we may generally expect a wider range of
factors in the gauge group since $f$ is more likely to contain
nontrivial monomials. Nevertheless, the
intermediate base generally admits further blow ups, and continuing
the blow-up process the whole sequence ends at a usual end point base,
such as the one with $h^{1,1}(B)=11341$.

The dynamic weight factor $D$ of such intermediate bases
$B_{\rm int}$ can be as
large as $\sim 10^{1,000}$ if we count $N_{\rm out}$ and $N_{\rm in}$
including all the possible blow up and blow downs
along the sequence leading to $B_{\rm int}$. However, this does
not mean that the actual number of these intermediate bases is
necessarily so large,
since we are tightly restricting the possible blow ups, and the actual
probability to get into such branches is exponentially small for any
given path (even
smaller than the  reciprocal of $D$, since the weight factor goes as
$\prod N_{\rm out}/N_{\rm in}$ while the probability for a given path
is $\prod 1/N_{\rm out}$).
Hence a given intermediate base reached by a given path does not really compete with the end
point bases in terms of total numbers.
It is rather unclear, however, whether in the overall graph the
intermediate good bases or the end point good bases dominate.  There
can be many ways in which the (4, 6) curves can be blown up from a
given point to reach an intermediate good base, and the number of
distinct topological types of intermediate good bases may be large
compared to the end points.  Further analysis is needed to determine
how these fit into the full distribution.  Note that in the analogous
situation of base surfaces, most of the good toric base surfaces are
not end points.

Another qualitative difference between the intermediate good bases and
the end point bases is the sensitivity to a small
perturbation. Suppose that we randomly blow down a base $B_{\rm int}$
two times and then randomly blow it back up two times.  Doing this
computational experiment, we find that in general we get a base with
the same $h^{1,1}(B)$ as $B_{\rm int}$, but with codimension-two (4,6)
loci. This indicates that the set of intermediate bases with a
particular $h^{1,1}(B_{\rm int})$ is sensitive to a small
perturbation. On the other hand, if we blow down an end point base
$B_{\rm end}$ two times and then randomly blow it up two times, we almost
always return to the exact same base that we started at (of course in
this random blow up process we only include the resolvable
bases). Usually we get the exact same base even if we blow it down and
then blow up $\sim$20 times. If we perform this process about
$100$--$500$ times, we generally get a
base that is similar to $B_{\rm end}$,
although transformed by a few flops, while the non-Higgsable cluster
structure remains the same. If we do this process $\sim 1000$ times,
we get another good end point base with a different non-Higgsable
cluster structure; for example one can simply check that the the number of
connected non-Higgsable clusters are different. 
More surprisingly, some bases are even more robust;
for
example if we take an end point base with $h^{1,1}(B)=2623$, then even
if we randomly blow it down 2400 times and then randomly blow up 2400
times, we still always get another end point base with the same set of rays as $B$ but different cone structure!
This other base actually shares the same set of toric rays as
$B_{\rm end}$, hence the base is only different up to a number of
flops. This clearly shows how robust these end point bases are, and
this observation and the large number of flops mentioned earlier
suggest that the good end points may actually dominate in the ensemble
of good bases.

\section{Conclusions and open questions}
\label{sec:con}

In this paper, following the earlier works \cite{MC, Halverson:2017},
we have explored the ``skeleton'' of the landscape of 4D ${\cal N} =
1$ F-theory vacua, which is a mathematically well-defined finite
graph whose
nodes are the toric threefolds that can act as bases of a Calabi-Yau
fourfold, and whose edges are given by toric blow-up and blow-down
transitions that take points and curves to divisors and vice versa.
Understanding the structure of this graph represents a first step
towards a global understanding of the ${\cal N} = 1$ landscape;
further steps would involve expanding the set of bases to include
singular and/or non-toric bases, considering the variety of elliptic
Calabi-Yau fourfolds over each base associated with Weierstrass
tunings, and including further information beyond the algebraic
geometry such as fluxes and T-brane information.

Even at the basic level of this ``skeleton,'' however, the structure
of this set of string compactifications is very rich.  This work and
the previous related works really have only begun to scratch the
surface of the set of questions that may usefully be asked of the
global structure of this space.  In this paper we have used a new
one-way Monte Carlo algorithm to investigate the global
distribution of bases and the features of ``good'' bases that have no
codimension-two (4, 6) curves.

One significant observation from the results of this paper and
\cite{Halverson:2017} is that when codimension-two (4, 6) curves are
allowed in the base threefold, bases with such loci dramatically
dominate the set of geometries.  Naively, the corresponding F-theory
models have massless strings associated with an infinite family of
massless excitations; an important question for understanding the
physics of 4D F-theory vacua is to characterize the physics associated
with these singularities in the elliptic fibration and to determine
the consequences of such curves for low-energy physics.

More generally, this work makes it clear that even the number of
distinct geometries in F-theory compactifications is an exponentially
large number, at least $10^{{3000}}$ by the underestimate of this
paper, when codimension-two (4, 6) curves are included, and likely much larger.  While it has been appreciated for some
years that the combinatorics of fluxes can generate such large numbers
of solutions in string geometries \cite{Grana:2005jc, Douglas:2006es,
  Denef-F-theory}, the fact that the number of distinct geometries
itself is already so large gives a new perspective on the string
landscape.  In particular, the tools of F-theory and the incredible
power of holomorphic structure give us a way of systematically
describing the connected space of all the toric threefold geometries,
describing the nonperturbative physics of the landscape globally
in a much more precise and controlled way than exists at present for
sets of flux vacua on distinct Calabi-Yau geometries
in other approaches to string compactification.

In this new Monte Carlo approach, beginning from various different
starting points, we have achieved an approximation of the structure of
the set of 3D toric bases without codimension-two (4,6) loci.  Many of
these bases are concentrated at discrete peaks in the range
$h^{1,1}(B)=1,000$---$20,000$; see Figure~\ref{f:good}.  The
significance of these classes of bases can be verified from the fact
that entirely different starting points lead to end points of the
one-way Monte Carlo trajectories with the same set of $h^{1,1}(B)$.
We have found some intriguing structure; in particular, the Hodge
numbers of the elliptic Calabi-Yau fourfolds over these good end point
bases are mirror to those of elliptic Calabi-Yau fourfolds over very
simple complex base threefolds.  Among other things, this suggests
that just as minimal bases that cannot be blown down have simple
structure (e.g. Mori theory), maximal bases that cannot be blown up
may have some kind of mirror of this simple structure.  More
generally, it would be very interesting to further investigate the
relations between the different bases within a single peak with the
same $h^{1,1}(B)$ in more detail. These bases seem to vary generally
have the same non-Higgsable gauge groups but not the same
non-Higgsable clusters, and it is not clear whether they will give
rise to isomorphic elliptic Calabi-Yau fourfolds or not. We will leave
these questions to further research.

These good end point bases seem to have a fairly universal
structure. The non-Higgsable gauge groups generally take the form
$E_8^a\times F_4^b\times G_2^c\times SU(2)^d\times G$ where $G$ is
another relatively small gauge group with a few factors such as SU(3)
or SO(8). The numbers of each of the primary factors in the gauge
group can be approximated by the empirical formula
(\ref{Ngaugegroups}). There is a significant peak $h^{1,1}(B)=2999$
with a single non-Higgsable SU(3) (and many non-Higgsable SU(2)
factors), which suggests potential phenomenological interest. However,
it is hard to tune a U(1) gauge group on this base due to the presence
of many $E_8$ gauge groups. So it is challenging to realize any
standard model-like field theory on it directly. Another interesting
phenomenon is that non-Higgsable $E_6$ and $E_7$ gauge groups are
clearly disfavored, at least among the end point good bases. If one
starts from a base $\tilde{\mathbb{F}}_{12}$ with a non-Higgsable
$E_7$, one still does not get any non-Higgsable $E_7$ gauge groups at
the end points.

Using a dynamical weighting factor in the one-way Monte Carlo program,
an (under) estimate
 of the total number of good bases without any codimension-two
(4,6) locus is on the order of $10^{250}$---$10^{270}$. This number
is significantly bigger than the previous estimation $10^{48}$ in
\cite{MC}. A lower limit of the total number of resolvable bases with
codimension-two (4,6) loci that
one can get from blowing up one weak Fano
base is at order of $10^{1,900}$---$10^{3,000}$ for different starting point bases,
which is significantly bigger than the lower bound $10^{755}$ in
\cite{Halverson:2017}. These numbers are  all
likely dramatically underestimated for reasons that we attribute
primarily to systematic errors from the
abundance of starting points that we have not systematically
classified but which feed into the graph in a way that suppresses the
weight factor, leading to an underestimate of the total number of
nodes in the graph.

In this paper, we also constructed two further special classes of
bases. The first one is the class of intermediate good bases that can
be generated by blowing up codimension-two (4,6) loci consecutively on
a resolvable base. The generic elliptic Calabi-Yau fourfolds over
these bases typically have a large $h^{3,1}$ on the order of
$10^2$---$10^3$. 
It is not clear whether the total number of bases in this class  is
much smaller than or greater than the total of end point bases,
although the stability of the end point bases under perturbation and
their large degeneracy under flops suggests that they may dominate. 
The second special class of bases we have discussed is the
set of exotic starting points constructed by blowing down the end
point bases randomly, which cannot be further blown down to another
smooth base. These exotic starting points are generally resolvable
bases with a large number of codimension-two (4,6) loci, but they are
still a crucial component in understanding the full structure of the
set of toric threefold bases in F-theory landscape, because of their
status as starting points before the blow ups. It would be interesting
to estimate their total number and study their structures, and/or to
extend the set of nodes of interest by including singularities in the base.

Of course, it is also very important to clarify the physical
consequences of not only codimension-two (4,6) loci, but also of
codimension-three (4, 6) loci and
terminal singularities. The resolution of a singular elliptic fourfold
with a codimension-two (4,6) locus or codimension-three (4,6) locus
involving an $E_7$ gauge group will lead to a
non-flat fibration. There
is a general statement that a tensionless string will arise in all
these cases
\cite{Braun:2013nqa} from M5 branes wrapping the 4-cycles in the
non-flat fiber. It will be very interesting if one can find an
$\mathcal{N}=1$ SCFT description. On the other hand, there is no
classification of general codimension-three (4,6) loci and terminal
singularities in the elliptic Calabi-Yau fourfold yet. We leave  the
further study of when a codimension-three (4,6) locus leads to non-flat
fibration or non-resolvable singularity to future work.

With such a huge data set of resolvable bases and good bases that can be
accessed on this ${\cal N} = 1$ ``skeleton'' graph using the
techniques of this paper and more generally by traversing the graph
using blow up and blow down transitions to move along the edges, and
in particular with such a large number of bases available and only
partial information so far about the global structure, it
may be interesting to extract more physical information
and patterns using machine learning methods. Such an
approach has already
been explored in \cite{Carifio:2017bov}, and we hope to learn more
about this data set using a variety of methods
in future work.

Finally, a complete understanding of the set of threefold bases used
in 4D F-theory compactification requires a classification of non-toric
threefold bases. A first step is to try to blow up a generic toric
base once on a non-toric curve or point. We will try to estimate the
total number of such blow ups and obtain a lower bound on the number
of non-toric bases in a future work. Of course, most of the bases we
get from this process are only resolvable, and it is very interesting
to ask whether and how these non-toric bases can converge to some
non-toric end points similar to the toric case by continuing the
blow-up process.

\section{Acknowledgements}

We would like to thank Thomas Grimm, James Halverson, Ling Lin, Cody
Long, and David Morrison for useful discussions. This research was
supported by the DOE under contract \#DE-SC00012567.

\appendix

\section{A toy model of one-way Monte Carlo with a lognormal
  distribution}

To illustrate the effectiveness of the lognormal distribution in
compensating for the difficulty in sampling the long tail in a random
walk on a highly branched layered graph, we consider a simple example.

Consider a graph containing $N$ nodes at each level $k$, where $N$ is
very large, with the following structure: Each node at level $k$ has
either one or two outgoing edges, each with probability 1/2.  Each
node at level $k + 1$ has either one or two incoming edges, again
with
probability 1/2.  (To be precise, we can impose the condition that
the outgoing and incoming branches are distributed
randomly, but subject to the constraint that the total number of edges
between level $k$ and level $k + 1$ is precisely $3 N/2$.)
The $3 N/2$ outgoing edges from level $k$ are
connected randomly to the incoming edges at level $k + 1$.

Now assume that we initiate a random walk at an arbitrary initial node
at level $k = 0$, with the weight factor $D = N$ in all cases. This
initiates the random walk with the condition that $\langle D \rangle =
N$ as desired.  Now, let us consider the possible changes that may
occur to the dynamical weight factor when performing a one-way Monte
Carlo on the graph, according to (\ref{dweight}), when going from
level $k$ to level $k + 1$.  We can have either
$N_{\rm out} = 1$ or $N_{\rm out} = 2$ each with probability 1/2.  And
we have
$N_{\rm in} = 1$ or $N_{\rm in} = 2$, but with probabilities 1/3 and
2/3 respectively.  Thus, we can have the following changes in the
dynamical weight factor:
\begin{equation}
D (k) =D \; \; \rightarrow \; \;
D (k + 1) =
\begin{array}{ll}
D/2, \;\;\;\;\;& {\rm probability} = 1/3,\\
D, \;\;\;\;\; &{\rm probability} = 1/2,\\
2D, \;\;\;\;\;& {\rm probability} = 1/6 \,.
\end{array}
\end{equation}
Note that the average value of $D(k +1)$ is $\langle D \rangle$ as
desired.  But over many iterations, this procedure will give a
lognormal distribution, with a very long tail.  A typical Monte Carlo
run will give a value of $D$ that diminishes exponentially with the
number of steps.  To analyze this explicitly, we consider the effect
of the one-way Monte Carlo step on the logarithm of $D$ base 2:
\begin{equation}
\log_2 D (k + 1) = \log_2D (k) + x \,,
\end{equation}
where $x$ is a random variable taking the value $-1$ with probability
1/3, $0$ with probability 1/2, and $+ 1$ with probability 1/6.  We can
easily compute
\begin{equation}
x_0 =\langle x \rangle = -1/6, \;\;\;\;\;
\sigma^2 =
\langle x^2 \rangle -\langle x \rangle^2 = 17/36.
\end{equation}
After $K$ rounds of the Monte Carlo process therefore, we will have
a distribution on $\log_2 (D (K)/D (0))$ that will be peaked around
$Kx_0 = -K/6$, with a standard deviation of $\sqrt{K} \sigma$.  For
example, for $K = 100$, a typical run will give a value like $D (K)
\cong 10^{-5} N$, even though the average over an exponentially large
number of runs will eventually give $\langle D \rangle = N$.

Given a finite sampling of runs of the one-way Monte Carlo algorithm
on this graph (i.e. a number of runs that is small compared to $e^K$),
the proper way to accurately estimate the total number of nodes at
level $K$ is to estimate
the mean and standard deviation of the logarithm $x$ as described
above.  We can then properly estimate the number of nodes at level $K$
assuming the lognormal distribution as
\begin{equation}
\langle D (K) \rangle \equiv
\int dx \; \frac{1}{ \sqrt{2 \pi} \sqrt{K} \sigma} 
e^{-(x-Kx_0)^2/2K \sigma^2} e^{x \ln 2}
= 2^{K (x_0 + \sigma^2 (\ln 2)/2)}
= 2^{K (-1/6 + 17 (\ln 2)/72)} \,.
\end{equation}
Evaluating numerically, we see that $17 (\ln 2)/72 \equiv 0.16365$, so
the correction factor
\begin{equation}
{\rm correction} = e^{\sigma^2 (\ln 2)^2/2}
\end{equation}
almost precisely captures the proper rate of growth (or absence
thereof in this case) of $D$ in the one-way Monte Carlo.

This method only works
in this form when the
distribution on $d_k = D (k + 1)/D (k)$
is independent of $k$ and uniform
across the graph, so that the
standard deviation scales as $\sigma\propto\sqrt{k}$.

Another way to estimate the standard deviation
$\sigma=\sqrt{\frac{17K}{36}}$, which is applicable in a broader range
of circumstances, is to compute the ``local'' standard
deviation of $f(k)=\log_2(d_k)=\log_2(N_{\rm out}(a_k)/N_{\rm
  in}(a_{k+1}))$: 
\be
\sigma(a_k)=\sqrt{\frac{\sum_{i=k-l}^{k+l}\left(f(i)-\frac{1}{2l+1}\sum_{j=k-l}^{k+l}f(j)\right)^2}{2l+1}}
\ee
for each node $a_k$. 
The $\sigma$ in the compensation formula (\ref{Expwf}) then becomes
\be
\sigma=\sqrt{\sum_{i=l+1}^{k-l}\sigma^2(a_i)}.\label{rsigma}
\ee
Because this method of computation is local in $k$ and only depends on
the region of the graph through which a single trajectory passes, it
may be useful in a much broader range of circumstances, in particular
for the one-way Monte Carlo algorithm of this paper where the
connectivity of the graph depends on the level $k$ and region in the
overall graph being traversed.  Nonetheless, this method can still be
used in the more specialized case  of the homogeneous toy model
described above.
For example, we can generate a random sequence of $f(k)$ in the toy model, with $k=1$---$1,000$ and $l=4$. A computation shows that the $\sigma$ for the largest possible $k=1,000$, (\ref{rsigma}) gives $\sigma=19.9276$, which is close to $\sigma=\sqrt{17\times 1000/36}=21.7307$ from the previous analytical computation.

\section{An example of
an exotic starting point}

Here we explicitly present the data of an exotic starting point
$B_{\rm ex}$ with $h^{1,1}(B_{\rm ex})=59$, which cannot be contracted
to another smooth base. We list the toric rays in Table~\ref{t:rays}
and the 3D cones in Table~\ref{t:cones}.

\begin{table}
\centering
\begin{tabular}{|c|c|c|c|c|c|}
\hline
$v_0$ & (0,0,1) &  $v_{21}$ & (3,4,1) & $v_{42}$ & (-3,-6,-2)\\
$v_1$ & (0,1,0) &  $v_{22}$ & (0,1,-1) & $v_{43}$ & (-11,-25,-8)\\
$v_2$ & (1,0,0) &  $v_{23}$ & (0,2,-1) &  $v_{44}$ & (-5,-6,-3)\\
$v_3$ & (-1,-1,-1) &  $v_{24}$ & (-6,-7,-4) &  $v_{45}$ & (-12,-29,-9)\\
$v_4$ & (1,1,1) &   $v_{25}$ & (1,1,0) &  $v_{46}$ & (-12,-15,-8)\\
$v_5$ & (-1,-1,0) &   $v_{26}$ & (1,-2,1) &  $v_{47}$ & (-7,-16,-5)\\
$v_6$ & (-2,-2,-1) &   $v_{27}$ & (-1,-4,-1) & $v_{48}$ & (2,-1,4)\\
$v_7$ & (0,-1,0) &   $v_{28}$ & (2,1,0) &    $v_{49}$ & (1,-1,4)\\
$v_8$ & (-1,-3,-1) &  $v_{29}$ & (-1,-5,-2) &  $v_{50}$ & (-13,-26,-9)\\
$v_9$ & (-1,0,-1) &  $v_{30}$ & (2,-1,2) &   $v_{51}$ & (1,-1,5)\\
$v_{10}$ & (-2,-5,-2) &  $v_{31}$ & (4,4,1) &  $v_{52}$ & (-2,-7,-2)\\
$v_{11}$ & (2,2,1) &   $v_{32}$ & (2,-1,3) &   $v_{53}$ & (-10,-20,-7)\\
$v_{12}$ & (0,4,-1) &  $v_{33}$ & (1,-1,2) &   $v_{54}$ & (-8,-19,-6)\\
$v_{13}$ & (1,-1,1) &  $v_{34}$ & (-7,-9,-5) &  $v_{55}$ & (-13,-25,-9)\\
$v_{14}$ & (-3,-3,-2) &  $v_{35}$ & (1,-1,3) &  $v_{56}$ & (-17,-35,-12)\\
$v_{15}$ & (-4,-5,-3) &  $v_{36}$ & (-1,-2,-1) & $v_{57}$ & (-4,-9,-3)\\
$v_{16}$ & (0,-5,-1) &  $v_{37}$ & (-10,-13,-7) &  $v_{58}$ & (-5,-13,-4)\\
$v_{17}$ & (-3,-5,-2) &  $v_{38}$ & (-4,-10,-3) &  $v_{59}$ & (-11,-24,-8)\\
$v_{18}$ & (2,3,1) & $v_{39}$ & (-3,-4,-2) &   $v_{60}$ & (-14,-29,-10)\\
$v_{19}$ & (2,-1,1) & $v_{40}$ & (-9,-11,-6) &  $v_{61}$ & (-15,-33,-11)\\
$v_{20}$ & (3,3,1) & $v_{41}$ & (-7,-15,-5) & & \\
\hline
\end{tabular}
\caption[E]{\footnotesize The list of toric rays of the exotic starting point $B_{\rm ex}$.}\label{t:rays}
\end{table}

\begin{table}
\centering
\begin{tabular}{|c|c|c|c|c|c|}
\hline
(22,23,9) & (13,26,19) & (19,30,2) & (19,30,13) & (30,32,2) & (30,32,13)\\
(34,40,24) & (34,40,37) & (37,46,40) & (32,48,2) & (32,48,35) & (49,51,26)\\
(49,51,49) & (53,55,50) & (53,56,50) & (55,60,59) & (60,61,59) & (52,27,16)\\
(28,2,23) & (54,53,56) & (51,48,2) & (33,32,13) & (33,13,26) & (35,32,33)\\
(35,33,26) & (49,48,35) & (49,35,26) & (20,2,28) & (20,28,31) & (28,21,31)\\
(2,16,12) & (28,21,25) & (28,25,23) & (41,45,43) & (59,53,55) & (59,53,54)\\
(1,9,23) & (1,23,25) & (21,31,20) & (38,45,43) & (58,27,52) & (27,38,45)\\
(21,20,18) & (2,34,24) & (36,34,37) & (27,59,54) & (27,54,45) & (2,22,23)\\
(1,21,25) & (7,19,2) & (41,45,54) & (41,54,56) & (6,40,24) & (46,44,37)\\
(2,24,15) & (24,14,15) & (58,55,57) & (2,11,20) & (11,19,20) & (46,44,40)\\
(58,55,60) & (42,41,56) & (42,56,50) & (58,60,61) & (58,61,59) & (58,59,27)\\
(7,38,43) & (2,9,22) & (14,15,2) & (7,47,43) & (17,42,50) & (8,12,16)\\
(17,57,55) & (17,55,50) & (7,26,19) & (3,14,9) & (39,36,37) & (6,44,40)\\
(2,9,3) & (0,2,51) & (6,29,44) & (2,3,14) & (7,27,16) & (7,16,2)\\
(18,21,1) & (39,29,36) & (8,2,12) & (39,37,44) & (30,44,29) & (7,27,38)\\
(10,29,6) & (6,14,24) & (36,29,2) & (36,2,34) & (2,4,11) & (8,10,29)\\
(8,29,2) & (8,16,52) & (6,14,9) & (0,26,51) & (7,0,26) & (1,6,9)\\
(8,17,52) & (17,52,57) & (52,58,57) & (42,47,7) & (42,47,43) & (42,43,41)\\
(4,1,18) & (4,18,11) & (4,2,0) & (8,10,17) & (10,6,17) & (7,6,5)\\
(7,5,0) & (7,42,6) & (42,17,6) & (1,6,5) & (1,5,0) & (1,0,4)\\
\hline
\end{tabular}
\caption[E]{\footnotesize The list of 3D fans of the exotic starting point $B_{\rm ex}$. We use $(i,j,k)$ to denote the triple element set $(v_i,v_j,v_k)$.}\label{t:cones}
\end{table}

If we try to contract a $\mathbb{P}^2$ divisor on $B_{\rm ex}$, for example the divisor corresponding to the ray $v_{61}$, then the volume of the new 3D cone $v_{58} v_{59} v_{60}$ is 2. Hence the resulting base after the contraction is singular. This is a well known feature of the toric Mori program, see \cite{ToricMori} for example.

\end{document}